# Statistics on 24 spiral galaxies having different observed arm locations using different arm tracers

Jacques P Vallée

National Research Council of Canada, Herzberg Astronomy & Astrophysics, 5071 West Saanich Road, Victoria, B.C., Canada V9E 2E7 jacques.vallee@nrc-cnrc.gc.ca

Abstract. The density wave theory predicted some physical offsets among different tracers of star formation. To test this prediction, here we compiled data on 40 galaxies searched observationally for a physical offset between spiral arm tracers, and found that 24 of them have a positive offset. In a spiral arm, an arm tracer in a region with a given temperature may be at a different location (offset) than an arm tracer in a region with a colder temperature.

Some conditions are found to be necessary or sufficient in order to detect an offset between two arm tracers. To find the offset of a tracer from another tracer, one needs a proper linear resolution. Starting in the dust lane and going across the spiral arm, we seek the observed physical width of the star-forming zone (offset). In our sample of 24 galaxies with measured offsets, we find offsets with a median value near 326 pc and a mean near 370 pc. These offsets are comparable to those found in our Milky Way galaxy, between the cold diffuse CO 1-0 gas set at 0 pc, and the hot dust near 350 pc.

Preliminary statistics are performed on the angular velocity of the gas and stars and angular velocity of the spiral pattern. Their observed orbital velocity of 200 km/s at a typical galactic radius near 4 kpc yields an angular speed of the gas and stars near 60 km/s/kpc. Their deduced angular rotation for the spiral pattern averages 36 km/s/kpc. These observational results are close to the results predicted by the shock-induced star-forming density wave theory. These mean or median property values will be useful for finding other galaxies that can support density waves.

## 1. Introduction

When one looks at a spiral arm with a given tracer (dust, hot gas, magnetic field, cold gas, etc), does one see an offset from what is seen with another tracer? The density wave theory predicts a shock when the orbiting gas reaches the spiral arm pattern, creating a dust lane on the arm's inner edge (in the direction to the Galactic Center). Observable chemical features were predicted upstream or downstream of the shock, notably infrared hot dust and radio masers upstream, but cold CO gas and old stars downstream, thus showing a "spatial ordering".

Such a "spatial ordering" could be searched as a "physical offset among different physical tracers" (dust, radio masers, infrared young stars, optically visible stars, HI, CO, old stars) across the width of a spiral arm. Some authors refer to this offset as the T-parameter, or the "time difference between two phases of star formation", or the 'time required to cross the observed angular offset" (Cedrés et al 2013).

Shu (2016) mentioned the apparent early failure of "not detecting color gradients associated with the migration of OB stars whose formation is triggered downstream from the spiral shock front". Such "color gradients" were often searched as "age gradients" in OB stars.

Most workers in the field no longer dispute that shock waves induced in the cold interstellar medium are behind this phenomenon; however, there is considerable controversy whether OB star formation is triggered by such shock waves, or whether OB star formation occurs throughout the disk of spiral galaxies, and whether the role of spiral density waves is simply to concentrate the molecular clouds in the arm regions by the mechanism of "orbit crowding." Since orbit crowding occurs even in the spiral shock calculations of W. W. Roberts and others, then subtle methods are required to distinguish between theories of "orbit crowding only" and "orbit crowding leading to shockwave triggering of OB star formation and subsequent aging of Population I objects downstream from the shock front identified by sharp and narrow dust lanes".

The competing dynamic spiral models can also provide some offsets, but with no systematic radial dependence (almost random offsets).

This paper is a compilation of spiral arm tracer offsets from the literature, providing the current observational situation in nearby spiral galaxies. There is a need for a meta-analysis of recently published data (last 12 years), to search for physical offsets. The various methods employed by others in this collection are different, as explained below, and a given method's bias is often countered by another method's bias, revealing reasonable statistical means, at the price of an enlarged broadness of the values.

All galaxies here are spiral galaxies, and all are nearby, probably formed in the same area (chemical composition, supercluster dynamics, etc). Some differences can be expected (flocculent versus grand design, sizes and masses, etc), yet we expect our broad statistical analysis to provide reasonable statistical means, at the price of an enlarged broadness of the values.

We then follow on the nature of the displacements observed in different spiral arm tracers (Figures 1 to 5), and the information that they contain concerning the difference speed between the spiral pattern and the material objects in the galaxy (Figures 6 to 10). A comparison with the Milky Way situation follows (Fig. 11).

Theoretical reviews on the subject have been published elsewhere (e.g., Dobbs et al 2011; Dobbs and Baba 2014; Dobbs et al 2015; Gittins & Clarke 2004; Roberts 1975; etc) and are briefly discussed here in conjunction with our statistical results (e.g. Fig.12).

Plan of this study. In Section 2, we list the primary data set, noting their averaging bin (resolution) employed. In Section 3, we discuss collectively a set of conditions to detect a physical offset. In Section 4, we provide a physical analysis and statistics for this data set. In Section 5, we provide a comparison of these first-order results for nearby spiral galaxies, with those for the Milky Way spiral galaxy. In Section 6, some comparisons are made with some theoretical predictions. A concluding discussion follows in Section 7. Some alternate theories are briefly discussed in an Appendix.

## 2. Observed data

Here we surveyed the literature over the last 12 years for many nearby spiral galaxies (located within 110 Mpc), having been looked at for the presence or absence of a physical offset.

These published results looked for a physical or an angular offset, in the *azimuthal direction along a circular orbit* around the galaxy's center.

Table 1 gives, for each attempt to measure an offset, the galaxy's name, distance, bin resolution used for the fitting function 'y' (often 'y' is the telescope resolution, or a smoothing or superpixellation), the two tracers compared, whether an offset was found or not, and the reference. Here the physical offset is between the two tracers employed.

Table 1 repertories 46 negative results and 45 positive or potentially positive results. The negative offset results may be due to taking two tracers from the same cold gas reservoir, or two tracers from the same warmer gas reservoir (too close in space). A larger discussion on this issue is provided below.

3. A collective discussion on yes and no results, for the same galaxy

Which fitting method or function 'y' ? Some potentially positive offset results came with the word 'ambiguous' in the reference cited, having a relatively large error. The tracer known as the Q-index method samples both the cold gas and the warmer gas, making it difficult to find a firm spatial boundary for a physical offset. Martinez-Garcia et al (2009a, 2009b) employed the Q index (a function of 4 wavelengths in the optical bands: g=0.50μm, r=0.6800μm, i=0.78μm, J=1.25μm) rising when there is a star-formation burst some 20 Myrs after a shock, then falling afterwards later in time (colder gas). Complex functions such as the Q function requiring observations at different wavelengths, quickly become model dependent and based on model assumptions.

Which coverage in 'x' ? If all the parameters were the same (resolution, tracers, etc), except for the fitting function y(x) and its covered range x (here 'y' can be a bye-eye fit or a polynomial fit of the peak, covering a short or long range in 'x'), then such choices can become subjective in some cases. For example, for each galaxy that Tamburro et al (2008) said yes, then Foyle et al (2011) said no (fig. 14 and section 4.3 in Foyle et al 2011). Later on, the results of Foyle et al (2011) were questioned by Martinez-Garcia & Puerari (2014), who employed a Fourier transform method ('y') to analyze the signal intensity over a 360$^o$ azimuthal range ('x'), at an angular resolution of 13"; they found an offset between CO and 24μm tracers for NGC 628 and NGC 5194. The HI tracer used in Foyle et al (2011) does not take account of the large-scale dissociation of molecular gas by newly formed stars in galaxies (Allen 1986).

Which two tracers? The offset searched or measured is the physical displacement between the two tracers. Two tracers that are too close in timescale may not yield a consistent offset (OB stars in B, V, I versus V band, say). Many observations were made with many star formation tracers at optical and near-infrared wavelengths. Some have argued the merit of some particular tracers over many other tracers or a combination of filters. Similarly, warmer dust (seen in the near infrared) traces a lane where some star formation is occurring and where the masers are located (as seen in the radio domain).

Which gas? HI gas may be found by photodissociation in the mid-arm owing to photons from old optically visible stars, and may be partially found in the compressed dust lane at the arm's inner edge. Louie et al (2013) have proposed that the choice of two starforming tracers (too close in time) may be the explanation of finding no offset, in some cases. In the shocked dust lane, the gas gets compressed and star formation can be initiated from the shocked condensed gas; hence the compressed CO gas seen there is shocked and hot within the lane width. But most observed diffuse CO 1-0 gas is colder and located elsewhere (near the location of the spiral 'potential minimum' – see Fig. 2 in Roberts 1975), seeable with a broad telescope beam as the observer moves the telescope in galactic longitude and notes the gas intensity peaking at specific longitudes (Tables 3 and 5 in Vallée 2016).

Which resolution? A sufficiently high angular resolution is obviously needed so that the physical offset is detected (avoiding beam dilution), but the bin size must not be too small so that the signal is loss below the instrumental noise (the beam sensitivity limit). Hence the small bin linear size is a necessity, but not a guarantee, of finding an offset.

Which galactic theory? Not all galaxies may have density waves. Some spiral galaxies may not possess a density wave pattern (q=0), or some may have multiple overlapping density wave patterns (q=2 or more; Meidt et al 2008), thus without a clear physical offset. Here the interpretation of a physical offset is based on the density-wave theory (q=1). Alternative explanations for a physical offset are mentioned below.

The detection of a physical offset among tracers requires attention to several conditions. Table 2 lists several conditions that should be necessary or sufficient to detect a physical offset. One would hope that adhering to these conditions would enable some more positive offsets to be found. The arguments in Table 2 are provided as a guidance for future observations; their relative importance still need to be assessed technically.

## 4. Physical analysis and statistics on the detected physical offsets between arm tracers.

Different methodologies were employed to find tracer offsets. Despite a wide inhomegeneity in looking for an offset, those detected can give us a first-order view of what they are. In turn, this first-order view is good enough to compare statistically nearby spiral galaxies, and compare that with the view for the Milky Way galaxy.

Table 3 assembles the positive physical offsets listed in Table 1. The physical offset is the displacement between the two tracers employed.

The statistics at the bottom of the table indicated median and mean values. Thus here for these galaxies the median physical offset is 326 pc, while it is 370 ± 50 pc for the mean value.

In practice, each method is complicated, and finding a mean correction for a single galaxy requires that galaxy to have been measured by all methods. At least 6 methods are listed in Table 4, while the one galaxy in Table 4 measured by the most methods is M51 with only 3 methods (out of 6). Also, the mean corrections for the interacting M51 galaxy may be different than a mean correction re-done for an isolated galaxy.

In addition, the bias of each method ('y') employed can statistically be counteracted by the bias of another method, thus allowing statistical means to be reasonably accurate for our first view, at the expanse of having a broader range ('x').

Figure 1 shows an histogram of all the physical offsets obtained – all data from Table 1. A ¾ majority of detections indicate a physical offset below 600 pc.

Figure 2 shows an histogram of the observed orbital velocity of the gas and stars around their galactic centers, at the galactic radius where the mean offset of tracers were observed. There is a median value of 200 km/s and a mean value of 198 ± 13 km/s for this orbital velocity (Table 3).

Figure 3 shows an histogram of the median galactic radius where the tracer offsets were measured. There is a median value of this galactic radius near 4 kpc (Table 3).

These statistics yield a consistent range of physical offsets, favoring the shocked-induced star formation in the dust lane located on the side of the spiral arm toward the galactic center, and the ensuing departure of the star-forming gas towards the middle of the spiral arm.

Next, we employed the density-wave theory to derive other galactic parameters. Where does starformation start? In this paper, we provide some numbers to some parameters, as did most papers tabulated in Table 1 (except for the Milky Way, and Chandar et al 2017). Alternative theories are discussed below.

In the density wave theory, the gas and stars move in a roughly circular orbit with a roughly constant velocity, while the density wave spiral pattern has a fixed angular rotation, so there is an offset between the two that is visible across a spiral arm. The shock occasioned by the arrival of a density wave produces a dark lane on one side of a spiral arm. There, new stars form and move out towards the middle of the arm, producing a physical offset between these two starforming phases.

Figure 4 shows the gas at the time of star formation (nascent star) in a dust lane (Red 1, at left), and its trajectory to show its position as an old star in the mid arm (Blue 2, at right).

The following equations describe the time elapsed between the two phases of star formation.

The angle is given by $A = R\ \Omega_{gas}\ T_{form}$
where the value $\Omega_{gas}$ is the circular gas velocity around the Galactic Center divided by its galactic radius R, and $T_{form}$ is the time elapsed between these two phases of star formation.

Similarly, the middle of the spiral arm moved in that time period, from Blue 1 at left, to reach Blue 2 at right, covering a somewhat smaller trajectory.

That angle is given by $C = R\ \Omega_{sp}\ T_{form} = A - B_{lin.offset}$
where the value $\Omega_{sp}$ is the angular speed of the spiral arm pattern around the Galactic Center, and $B_{lin.offset}$ is the observed physical offset between the dust lane Red 2 and the mid-arm Blue 2.

At a given galactic radius R, the rotation speed of the gas and stars ($R\ \Omega_{gas}$) must be subtracted from the spiral pattern speed ($R\ \Omega_{sp}$) to yield the differential speed between the two, which can be multiplied by the star-formation timescale $T_{form}$ in order to yield the physical linear offset $B_{lin.offset}$ between the dust lane (inner arm edge) and the 'potential minimum' of the density wave where observationally the diffuse colder CO 1-0 sits (it peaks in intensity using a broad telescope beam size while scanning along the galactic longitudes).

Subtracting these equations, one gets

$$[R\ \Omega_{gas} - R\ \Omega_{sp}]\ T_{form} = k\ B_{lin.offset} \qquad \text{(equ. 1)}$$

where $B_{lin.offset}$ is observed, and so $T_{form}$ can be deduced (see Equ. 4 in Egusa et al 2009). Here R is in kpc, $\Omega$ is in km/s/kpc, $T_{form}$ is in Myr, and the physical offset $B_{lin.offset}$ is in parsecs (Vallée 2017a – his table 2). The coefficient k is 1 when using standard MKSA system units. When using non-standard system units, such as R in kpc, $\Omega_{sp}$ in km/s/kpc, T in Myr, B in pc, then k becomes $1.023 = 3.16\times10^{13}$s (in 1 Myr) / $3.09\times10^{13}$ km (in 1 pc).

It is noted that $\Omega_{gas}$ can be a function of the galactic radius R. Equation 1 allows the conversion of a physical offset (in parsecs) into a timescale (in millions of years).

Given that the parameters $\Omega_{gas}$, R, and $B_{lin.offset}$ can be observed at a telescope, then it follows from this equation 1 that $T_{form}$ and $\Omega_{sp}$ are now related. Since $\Omega_{sp}$ cannot be directly determined from observations, one must employ indirect methods and assumptions, giving rise to a larger error bar. The estimates of $\Omega_{sp}$ are still rather ambiguous, leading to uncertainties. It is likely that $T_{form}$ can be dependent on the galactic radius, due to the metallicity gradient in disk galaxies.

Table 4 assembles these parameters, pertaining to the angular speed of the spiral pattern, and to the star formation time scale.

Figure 5 shows an histogram of the deduced starforming timescale (from new stars to old stars), in a spiral arm. There is a median value of 4 Myr and a mean value of 7.4 ± 1.5 Myr (Table 4).

Figure 6 shows an histogram of the deduced angular spiral pattern speed around their galactic centers. There is a median value of 31 km/s/kpc and a median value of 36 ± 6 km/s/kpc (Table 4).

Also, we have computed other physical parameters of note, such as the difference between the observed and predicted angular and linear velocities, for each spiral galaxy.

Table 5 shows these parameter values for the angular velocity of the gas and stars (from Table 3), the difference of that with the deduced angular pattern speed (Table 4), and the differential linear speed at the appropriate galactic radius (the shock velocity at the entrance of the spiral arm). When a galaxy has 2 or more entries, then we averaged here the various entries before proceeding.

$v_{gas} - v_{sp}$ is the difference in linear speed, with the spiral pattern speed at a given galactic radius r given by $v_{sp} = r\ \Omega_{sp}$ $R_{coro}$ is the co-rotation radius, defined as where the speed of gas and stars equals $v_{sp}$ .

Figure 7 shows an histogram of the angular velocity of the gas and stars. A median of 60 and a mean of 63 ±7 km/s/kpc are computed (Table 5).

Figure 8 shows an histogram of the difference in angular velocity, that of the gas and stars minus that of the spiral pattern, Positive values imply that a shock will occur near the entrance to the arm (inside edge). A median of 26 and a mean of 26 ±3 km/s/kpc are computed (Table 5).

Figure 9 shows an histogram of the shock speed (the difference of the speed of the gas and stars minus the speed of the spiral pattern). A median of 77 and a mean of 82 ±7 km/s are computed (Table 5).

Figure 10 shows an histogram of the co-rotation radius (the ratio of the circular orbital speed from Table 3 and the angular speed of the spiral pattern from Table 4). A median of 7 and a mean of 7.0 ±1 kpc are computed (Table 5).

It is interesting that the parameters so computed (shock speed, co-rotation radius, etc) give values consistent with the ranges and limits as predicted in the density-wave theory. The galactic radius where the parameters are computed is lower than the co-rotation radius, and the orbital linear speed of gas and stars is higher than that of the spiral pattern speed (enough to create a shock).

Discussion. Most observations in the literature of physical offsets (copied or derived anew in Table 1 and 3) assumed a nearly circular velocity to derive certain quantities. Model with slightly elliptical orbits were discussed in Louie et al (2013), including wiggles (about 200 pc) that go back and forth in spiral arms that tend to cancel out over a full orbit, not biasing the statistical values. The long-term spiral arms, assumed in the density-wave theory, would require gas and young stars to have at most a small orbital eccentricity (Fig. 3 in Roberts 1975), as otherwise their orbits may have encounters at widely different galactic radii and be changed forever (not rejoining again the assumed long-term arms).

As seen in Tables 1 and 3, for each method, one of the tracer is selected to detect star formation. Which tracer best detect star formation is not a statistical issue, as the difference in offset is small compared to the arm's width – see Figure 11 for the Milky Way.

## 5. Comparison with earlier findings on physical offsets

Notable observed offsets are around 100 to 350 pc in the Milky Way (Vallée 2014a, 2014b), and around 200 pc in M51 (Chandar et al 2017).

Thus in the M51 spiral galaxy, Chandar et al (2017) found a typical physical offset near 200 pc, at a mean galactic radius between 3 and 6 kpc; this mean value covers several measurements, including a physical offset of 50 pc between dust lane and 3.6μm emission, then a 2nd offset of 220 pc between the dust lane and the bright emission from massive stars in stellar clusters, and a third offset of 140 pc between dust lane and infrared emission from old stars.

What is the true angular spiral pattern speed? For M51, the timescale varies from 3.4 to 13.8 Myrs (a ratio of 4.1), while the angular spiral pattern speed varies (from 21 to 40 km/s/kpc). As these two parameters (timescale $T_{form}$ and angular spiral pattern speed $\Omega_{sp}$ ) are related from equation 1 above, and the other parameters can be observed at a telescope, it follows that a large discrepancy in the angular spiral pattern speed will entail a large discrepancy in the starforming timescale. That large uncertainty may be due in part to the non-circular orbit.

In the Milky Way spiral galaxy, Vallée (2014) used the *arm tangents* to the spiral arms seen edge-on in galactic longitude, to locate the location of different star formation tracers, and found angular offsets among them.

Figure 11 shows the observed offsets for the crosscut of the Milky Way's spiral arms, as adapted from fig. 4 in Vallée (2017b). The direction of the Galactic Center is to the right. At left, the outer arm edge is shown, but there no star formation tracers there – nothing but widely distributed cold diffuse CO 1-0 gas and old stars. The peak of the cold diffuse CO gas marks the ‘potential minimum’ of the density wave, while the peak of the old stars is within its error (horizonthal blue dashes).

Color coding in this figure summarizes four zones, starting with a blue (cold) zone at the arm center where diffuse cold CO 1-0 occupies a large arm width, along with old optically visible stars, and HI gas, as well as old HII regions (near the ‘potential minimum’ of the density wave). This zone is followed by a green zone, about 100 pc away from the arm center, towards the Galactic Center, encompassing $^{13}$CO 1-0 molecules, relativistic synchrotron emission, and radio recombination lines from young HII regions. An orange zone follows, about 200 pc from the mid-arm, toward the Galactic Center, containing very young star forming regions, dense small CO clumps, FIR [CII] and [NII] emissions, and radio masers. Finally, a red (hot) zone follows at the arm’s inner edge, about 300 pc from the arm’s center, facing the direction to the Galactic Center, encompassing NIR and MIR emissions from shocked dust (near the shocked lane of the density wave).

In the Milky Way galaxy, Vallée (2016) found a mean typical physical offset near 350 pc, for a galactic radius between 4 and 5 kpc; this mean value covers several measurements, including an offset of 340pc between hot NIR dust and CO in Galactic Quadrant I, and an offset of 380 pc in Galactic Quadrant IV, as well as an offset from CO of 192 pc between radio masers in GQ IV compared to an offset of 210 pc in GQ I.

For the Milky Way, a similar spread in angular spiral pattern speed was noted (from 12 to to 30 km/s/kpc – see Vallée 2018), although recent observations in the location of masers on the inner Perseus arm (forcing the co-rotation radius to be beyond the Perseus arm) has strongly curtailed the range of the angular spiral pattern speed to be less than 21 km/s/kpc (Vallée 2018; Vallée 2019).

A similar figure has been constructed separately for the arms in Galactic Quadrant IV, and for the arms in Galactic Quadrant I, showing the reversal of the dust-CO pattern across the galactic longitude of the Galactic Center (Fig. 2 in Vallée 2016); the same half-arm width (from hot dust at the arm’s inner edge to the CO at the arm center) is found in Galactic Quadrant I (about 340 ±30 pc ), and found in Galactic Quadrant IV (360 ±30 pc).

Furthermore, a figure similar here has been constructed separately for two alternating sets of arms (Fig. 3 in Vallée 2014), showing the same arm half-width (set A: Sagittarius, Carina, Norma; set B: Scutum, Crux, Perseus origin), within their errors.

The bottom of Figure 11 shows the many long-sought gradients mentioned earlier (Section 1).

Appropriate tracers. What are the most appropriate arm tracers, to trace star formation over time? In the Milky Way galaxy, an appropriate tracer of a detected nascent star formation would be in the orange zone (peak of radio masers, say), while an appropriate tracer of old stars would be in the blue zone (peak of the broad diffuse radio CO 1-0, say). In the Milky Way, such a physical offset is near 350 pc.

Thus, first-order views of tracer offsets for some nearby spiral galaxies (Fig. 1) indicate a rough similarity with the offsets found for the Milky Way galaxy (Fig. 11).

### 5.1 Others

Others found a variable arm pitch angle value, as seen at different wavelengths (Pour-Imani et al 2016; Yu et al 2018; Miller et al 2019, and others). They have looked at an offset in an arm’s pitch angle, in the radial direction as measured from close to the galactic center and going outward in galactic radius.

Thus Pour-Imani et al (2016) looked at 41 spiral galaxies, measured the galaxy-wide pitch angle of its arms, and obtained that this galaxy-wide pitch angle of its arms is smaller at near-infrared (3.6μm; median 12$^{o}$) and optical B-Band (0.445μm; median 16$^{o}$) than at mid-infrared (8.0μm; median 21$^{o}$) and far ultraviolet (0.1516 μm; median 23$^{o}$). They offer possible interpretations on which arm tracers are predominant at each of these four wavelengths.

Yu et al (2018) measured via Fourier Transform the pitch angle for 113 galaxies at the 3.6μm, I, R, V, B, NUV and FUV wavelengths, and claimed that the pitch angle is smaller in the blue than in the I band, but found no difference in pitch angle between the 3.6μm and the I band.

Miller et al (2019) looked at 29 galaxies for their pitch angle value at different wavelengths. Their starforming tracers yielded a larger pitch angle value than their other tracers (older stars).

6. Limited comparisons with some theories predicting an offset

Here we do a quick comparison of the observational data with some theories. We do not present here a review of all the theories.

The presence of physical offsets with different arm tracers is a prediction of density-wave theory, and the predicted physical offsets are quite close.

The density wave theory with a shock-induced dark lane at the inner edge of a spiral arms predicted the physical width (offset) between the shock (at inner arm edge) and the spiral 'potential minimum' (where starformation ends), as a function of the orbital distance to the next arm – see Lin & Shu (1964), Roberts (1975 – his Fig. 2), and Gittins & Clarke (2004). The spiral 'potential minimum' location depends on many factors, such as on shock conditions (gas velocity, stellar mass distribution, etc – see Gittins & Clarke 2004 – their fig. 16, with P-arm and D-arm, versus SF-arm).

The density-wave theory discussed here made its predictions while employing a galactic radius between 8 and 10 kpc, while the observational data discussed here measured a detectable signal within a galactic radius between 2 and 6 kpc. Thus a re-scaling of the theoretical predictions is needed for a galactic radius nearer 4 kpc, before comparisons are done.

Fig. 2 in Roberts (1975) shows the shock location differs from the location of the 'potential minimum' by about *3.7 %* of the distance between the 'potential minimum' and the one in the next arm. For a galactic radius of 8 kpc, the circumference is 50.2 kpc, and each of 4 arms is separated by 12.6 kpc, Thus the small distance from the shock to the 'potential minimum' is 3.7% of 12.6 kpc, giving 0.5 kpc. All the tracer offsets in Figure 11 for the Milky Way spread over a maximum of 360 pc, thus between the shock location and the location of the 'potential minimum'.

Further out, his Figure 2 shows that the shock location differs from the location of the 'potential maximum' (middle distance to the next arm) by about 53% of the distance between the 'potential minimum' and the one in the next arm. Thus for the Milky Way and 4 arms, the large distance from the shock to the potential maximum is 53% of 12.6 kpc, giving 6.8 kpc; this would be located in the 'interarm' region.

Here we seek the offset predicted by other theories, between the shock location and the location of their 'potential minimum' (the rough end of the starforming phase).

In Fig. 3 of Tosa (1973), the shock is separated from the 'potential minimum' (at phase: zero) by *6.4%* of the full 360$^{o}$, computed at a galactic radius of 10 kpc and with 2 arms. For a galactic radius of 4 kpc, and a 4-arm model, the distance between the arms is 6.3 kpc, so a separation of 6.4% gives 396 pc.

In Figure 1 of Wielen (1979), one sees that the shock location differs from the distance to the 'potential minimum' by about 7.8 % of the distance between the 'potential minimum' (at phase 0$^{o}$) and the one in the next arm (at phase 360$^{o}$). His model was constructed at a galactic radius of 10 kpc. Rescaling it for a galactic radius of 4 kpc and 4 arms, the distance between each arm is 6.3 kpc, so 7.8% of that gives a separation as 490 pc.

In Figure 11 of Gittins and Clarke (2004), for 4 arms at a galactic radius of 10 kpc, the shock location ($\eta$= - 0.2) differs from the distance to the 'potential minimum' (at coordinate $\eta$ = 0) over a period of $2\pi$ (until the next arm). At a galactic radius of 4 kpc, their offset is 41% of their interarm distance (across the arm; their Fig.13), or 290pc for an interarm of 700 pc (their Fig. 15).

Finally, in Fig 4a of Dobbs & Pringle (2010), for a 2-arm galaxy with a density wave, an offset of $5^o$ (out of $180^o$, or 2.8%) at a radius of 7.5 kpc is predicted between young stars and stars of age 50 Myrs, or 650 pc. Rescaling it for a galactic radius of 4 kpc and 4 arms gives a separation as 176 pc.

Figure 12 here shows the observed values for the physical offset, and the predicted values for 4-arms models and separately for 2-arms models. These predicted values at a galactic radius of 4 kpc are similar to those observed in nearby galaxies at about the same radius (Table 3 here). Such a 4-kpc radius ensures that the co-rotation radius is farther out in these galaxies.

Such a reasonable agreement between theoretical and observed offsets point to the need for more searches for offsets, paying attention to the 10 conditions listed in Table 2. This could enhance the number of positive results (currently around half in Table 1).

Baba et al (2015) argued that the offsets depends on the galactic radius. There is certainly no offset predicted at the co-rotation radius. Here the observations are limited by detector sensitivities, which limit the range in galactic radius where the offset is measured.

## 7. Conclusion

In this work, we have investigated if some nearby galaxies showed a separation between one or more tracers of starformation (dust lane, masers, Hα lines, etc) and one or more tracers of normal or older stars (V-band image, old star clusters, CO 1-0 diffuse gas, etc).

From a search of the literature (NASA ADS, etc) covering the last 12 years, we obtained 40 nearby spiral galaxies with one or more published studies looking for a possible physical offset $B_{lin.offset}$ among tracers (Table 1). We listed possible reasons for a non-detection (Table 2).

Also, we compiled the physical parameters of the 24 spiral galaxies showing a physical offset between some tracers (Tables 3, 4, 5). These physical parameter values should help forthcoming theories or further amendments to existing theories.

In this paper, we have shown the following:

1- There is a presence of an observed physical offset among some spiral arm tracers. Some 24 galaxies have a detected physical offset (Table 1; Table 3), out of 40 galaxies listed, or 60%. It is possible that the choice of tracers selected hindered a detection, and that this ratio could be much higher when using other tracers (see Table 2).

2 - Most detected tracer offsets are observed close to the dust lane, within 600 pc (Fig. 1). Several conditions must be there, in order to detect an offset (Table 2).

3- When an offset is detected, most of these galaxies have some physical parameters comparable to those found in the Milky Way (Fig. 2, 3, 5 to 11).

4- The range of detected physical offsets are comparable to the range of predicted physical offsets from density-wave theories (Fig. 12).

5- Assuming that the nearby galaxies in Table 1 where chosen at random, it follows that at least 60% of them have a physical offset in between tracers, as predicted by the density wave theory. This number could be much higher depending on the conditions employed (Table 2).

6. For a typical nearby spiral galaxy, obeying the prescriptions of the density-wave theory, its observed mean physical parameters would be the mean values in Tables 3 to 5: mean offset of 370 pc, mean orbital velocity of around 200 km/s, at a mean galactic radius near 4 kpc, mean angular rotation of gas and stars near 63 km/s/kpc, mean deduced angular rotation of spiral pattern near 36 km/s/kpc, mean (shock) velocity differential of 77 km/s, and mean starforming timescale near 7 Myr.

Acknowledgements.

The figure production made use of the PGPLOT software at NRC Canada in Victoria. I thank a referee for useful, careful, insightful and historical suggestions, and a scientific editor for clarity.

## Appendix.

A1 - Starforming timescales - discussion

Non circular orbital gas motions could increase the timescale by a factor of 2 or so (Louie et al 2013 – their fig. 10), while it could be substantially less than a factor of 1.5 (Tamburro et al 2008 – their fig. 10). Correcting our data for the non-circular motion model of Louie et al (2013) by a factor of 2, then our mean star forming timescale becomes near 14 Myrs.

Some time must have elapsed between an invisible gas clump's collapse in the compressed hot dust zone (red) in Figure 11, and the detectable radio masers in the orange zone. The time elapsed between the red and orange zones should be added to the time from the detection of a detected nascent star (orange zone) to the star's old age (blue zone). Different mechanisms for star formation differ in their theoretical predictions. The presence of magnetic fields gave timescales near 5-10 Myrs, while the presence of supersonic turbulences may shorten it to 3 Myrs (Martinez-Garcia et al 2009a, 2009b).

A collapsing molecular clump (seen at radio) may well start star formation inside, and then could disperse the remaining fragments of molecular gas just outside, thus preventing the collapse of the remaining gas outside of the clump. This local gas dispersion would allow the overall cloud (around this clump) to have a longer cloud lifetime. Cloud lifetimes have been predicted and observed to last around 20 Myrs (Tamburro et al 2008 – their Section 1), even reaching to 100 Myrs (Elmegreen 2011).

A2 – Some alternative theories.

There are very few theories to create an offset between two tracers. In galaxies where no offset has been found, other models have been suggested (tidal arms, flocculent arms, etc).

Thus among alternate theories for the formation of spiral arms, Dobbs & Pringle (2010) proposed *tidally-induced* arms (their Fig. 4d) in a galaxy subject to a strong external tidal interaction, and flocculence-induced arms (their Fig. 4c) driven by local gravitational instabilities; in both cases a clear physical offset is not expected. In the 'dynamic transient recurrent spiral' model, Fig. 5 in Baba et al (2016) shows a spiral arm maintained at the edge where 2 tangential *epicyclic flows* collide, but there is almost no gas streaming through the stellar arm, and there is no systematic offset between the gas and the stars. The *collision and merging* of a small dwarf galaxy with the Milky Way was shown numerically to induce the formation of small spiral armlets in the bigger galaxy, but leaving no consistent offset between the gas and the stars in the arms (Fig. 11 in Struck et al 2011; Fig. 4 in Purcell et al 2011).

Some theories assert that there is no triggering of star formation in a spiral arm, or that clouds go through arms without much effect from the arm (Dobbs et al 2011; Dobbs et al 2015; Dobbs & Pringle 2009; Koda et al 2009; Baba et al 2017). Some theories do not point to the exact place where starformation occurs. It could be that starformation starts in giant molecular clouds, and that the cloud produces stars near the spiral arm center, and in the compressed dust lane at the entrance to the spiral arm. However, in our Galaxy, most masers are observed to be located very near the dust lane at the inner edge of spiral arms, and not far away (e.g., Vallée 2018 – his fig.1; Vallée 2016 – his fig.2). Possible triggering inside dust lanes are known (e.g., Elmegreen et al 2018).

Some theories assert that the orbital motion of clouds and gas, around the Galactic Center, would enter a spiral arm, then would 'stream' parallel and inside the arm for a while, then would exit the arm and carry on in its orbit around the Galactic Center. This would thus take a longer time to reach from one arm tracer to the next, inside the arm, even though the arm is thin.

The calculations of Roberts (1975) already showed that the gas orbits are not quite circular. After going through a spiral shock, gas and young stars flow to some extent along an arm, then flow to the interarm (Dobbs & Pringle 2010), Assuming a strict circular motion will affect slightly the determination of the spiral pattern speed for different galactic radii, but not enough to prevent the star-formation offset (Martinez-Garcia et al 2009b). An approximate correction for the radial orbital component was given by Grosbol & Dottori (2009, their equation 1, involving filters at J, H, and K wavelengths, as well as the approximate pitch angle), giving new answers that are slightly lower than previously published answers.

For the Milky Way, the Sun's motion beyond the motion of the Local Standard of Rest (LSR) is about 15 km/s, while that of the LSR's motion is close to 230 km/s, hence this correction is only about 7%, and thus the Sun's orbit is likely to oscillate back and forth around the mean orbit of the LSR. In doing statistics over two dozen nearby spiral galaxies, the statistical median and mean values ought not to move beyond their respective errors, for such small corrections (<10%).

**Table 1. Recent attempts (last 12 years or so) to detect physical offsets between arm tracers (including non-detections)**

| Galaxy name | Dist (Mpc) | resolution (pc) | Tracer1 | Tracer2 | Offset (yes or no) | Reference |
|---|---|---|---|---|---|---|
| NGC0224 = M31 | 0.8 | 16 | CO 1-0 | NearUV | no | Tenjes et al (2017) |
| NGC0578 | 22.7 | 440 | Q low | Q high | yes | Martinez-Garcia et al (2009a) |
| NGC0628 = M74 | 7.3 | 200 | HI | 24μm | yes | Tamburro et al (2008) |
| - | - | 230 | CO 1-0 | Hα | yes | Egusa et al (2009) |
| - | - | 200 | HI | 24μm | no | Foyle et al (2011) |
| - | - | 880 | 3.6 μm | 8μm | no | Kendall et al (2011;2015) |
| - | - | 1500 | I band | Hα | yes | Cedrés et al (2013) |
| - | - | 460 | CO 1-0 | 24μm | yes | Martinez-G. & Puerari (2014) |
| - | - | 440 | old clust. | young clust. | no | Shabani et al (2018) |
| - | - | 50 | CO 2-1 | Hα | yes | Kreckel et al (2018) |
| NGC0918 | 21.7 | 420 | Qlow | Qhigh | no | Martinez-Garcia et al (2009a) |
| NGC0925 | 9.2 | 267 | HI | 24μm | yes | Tamburro et al (2008) |
| NGC1417 | 57.1 | 1100 | Qlow | Qhigh | yes | Martinez-Garcia et al (2009a) |
| NGC1421 | 29.3 | 568 | Qlow | Qhigh | no | Martinez-Garcia et al (2009a) |
| NGC1566 | 18.0 | 880 | 3.6 μm | 8μm | no | Kendall et al (2011; 2015) |
| - | - | 440 | B band | old clust. | yes | Shabani et al (2018) |
| NGC2403 | 3.2 | 94 | HI | 24μm | yes | Tamburro et al (2008) |
| - | - | 94 | HI | 24μm | no | Foyle et al (2011) |
| - | - | 350 | 3.6 μm | 8μm | no | Kendall et al (2011;2015) |
| NGC2841 | 14.1 | 410 | HI | 24μm | yes | Tamburro et al (2008) |
| - | - | 410 | HI | 24μm | no | Foyle et al (2011) |
| - | - | 1400 | 3.6 μm | 8μm | no | Kendall et al (2011;2015) |
| NGC3031 = M81 | 3.6 | 105 | HI | 24μm | yes | Tamburro et al (2008) |
| - | - | 105 | HI | 24μm | no | Foyle et al (2011) |
| - | - | 88 | 3.6 μm | 8μm | yes | Kendall et al (2008) |
| - | - | 280 | HI | star fields | no | Choi et al (2015) |
| NGC3162 | 23.7 | 460 | Qlow | Qhigh | no | Martinez-Garcia et al (2009a) |
| NGC3184 | 11.1 | 323 | HI | 24μm | yes | Tamburro et al (2008) |
| - | - | 250 | CO 1-0 | Hα | yes | Egusa et al (2009) |
| NGC3198 | 14.0 | 880 | 3.6 µm | 8µm | yes | Kendall et al (2011;2015) |
| NGC3351 | 9.3 | 270 | HI | 24µm | yes | Tamburro et al (2008) |
| - | - | 270 | HI | 24µm | no | Foyle et al (2011) |
| NGC3521 | 10.0 | 291 | HI | 24µm | yes | Tamburro et al (2008) |
| - | - | 291 | HI | 24µm | no | Foyle et al (2011) |
| NGC3621 | 6.6 | 192 | HI | 24µm | yes | Tamburro et al (2008) |
| - | - | 192 | HI | 24µm | no | Foyle et al (2011) |
| NGC3627 = M66 | 9.2 | 269 | HI | 24µm | yes | Tamburro et al (2008) |
| - | - | 269 | HI | 24µm | no | Foyle et al (2011) |
| - | - | 583 | CO 1-0 | 24µm | no | Martinez-G. & Puerari (2014) |
| NGC3938 | 15.8 | 306 | Qlow | Qhigh | no | Martinez-Garcia et al (2009a) |
| - | - | 490 | CO 1-0 | Hα | yes | Egusa et al (2009) |
| - | - | 880 | 3.6 µm | 8µm | yes | Kendall et al (2011;2015) |
| NGC4254 = M99 | 16.5 | 320 | Qlow | Qhigh | yes | Martinez-Garcia et al (2009a) |
| - | - | 370 | CO 1-0 | Hα | yes | Egusa et al (2009) |
| NGC4303 = M61 | 16.0 | 570 | CO 1-0 | Hα | yes | Egusa et al (2009) |
| NGC4321 = M100 | 16.0 | 560 | CO 1-0 | Hα | yes | Egusa et al (2009) |
| - | - | 880 | 3.6 µm | 8µm | no | Kendall et al (2011;2015) |
| - | - | 194 | opt. | UV | no | Ferreras et al (2012) |
| NGC4535 | 16.1 | 570 | CO 1-0 | Hα | no | Egusa et al (2009) |
| NGC4579 = M58 | 18.0 | 880 | 3.6 µm | 8µm | yes | Kendall et al (2011;2015) |
| NGC4736 | 5.1 | 170 | CO 1-0 | Hα | yes | Egusa et al (2009) |
| NGC4939 | 46.5 | 901 | Qlow | Qhigh | no | Martinez-Garcia et al (2009a) |
| NGC5055 = M63 | 7.8 | 227 | HI | 24µm | yes | Tamburro et al (2008) |
| - | - | 227 | HI | 24µm | no | Foyle et al (2011) |
| NGC5194 = M51 | 7.8 | 260 | CO 1-0 | Hα | yes | Rand & Kulkarni (1990) |
| - | - | 226 | HI | 24µm | yes | Tamburro et al (2008) |
| - | - | 270 | CO 1-0 | Hα | yes | Egusa et al (2009) |
| - | - | 940 | B band | B, V, I | no | Scheepmaker et al (2009) |
| - | - | 5 | V band | O,B assoc | no | Kaleida & Scowen (2010) |
| - | - | 226 | HI | 24µm | no | Foyle et al (2011) |
| - | - | 880 | 3.6 µm | 8µm | no | Kendall et al (2011;2015) |
| - | - | 160 | CO 1-0 | Hα | yes | Louie et al (2013) |
| - | - | 490 | CO 1-0 | 24µm | yes | Martinez-G. & Puerari (2014) |
| - | - | 76 | CO 1-0 | 3.6µm | yes | Chandar et al (2017) |
| - | - | 240 | HI | opt. stars | yes | Egusa et al (2017) |
| - | - | 37 | CO 1-0 | Hα, 24µm, B | no | Schinnerer et al (2017) |
| - | - | 440 | old clust | young clust | no | Shabani et al (2018) |
| NGC5248 | 22.7 | 760 | CO 1-0 | Hα | no | Egusa et al (2009) |
| NGC5371 | 43.5 | 843 | Qlow | Qhigh | no | Martinez-Garcia et al (2009a) |
| NGC5457 = M101 | 7.2 | 200 | CO 1-0 | Hα | yes | Egusa et al (2009) |
| - | - | 1466 | I band | Hα | yes | Cedrés et al (2013) |
| NGC6181 | 37.0 | 600 | CO 1-0 | Hα | no | Egusa et al (2009) |
| NGC6946 | 5.5 | 160 | HI | 24µm | yes | Tamburro et al (2008) |
| - | - | 160 | CO 1-0 | Hα | yes | Egusa et al (2009) |
| - | - | 160 | HI | 24µm | no | Foyle et al (2011) |
| - | - | 880 | 3.6 µm | 8µm | no | Kendall et al (2011;2015) |
| - | - | 1100 | I band | Hα | no | Cedrés et al (2013) |
| NGC6951 | 24.9 | 483 | Qlow | Qhigh | no | Martinez-Garcia et al (2009a) |
| NGC7125 | 44.6 | 864 | Qlow | Qhigh | no | Martinez-Garcia et al (2009a) |
| NGC7126 | 44.6 | 864 | Qlow | Qhigh | no | Martinez-Garcia et al (2009a) |
| NGC7753 | 72.1 | 1400 | Qlow | Qhigh | no | Martinez-Garcia et al (2009a) |
| NGC7793 | 3.8 | 112 | HI | 24µm | yes | Tamburro et al (2008) |
| - | - | 112 | HI | 24µm | no | Foyle et al (2011) |
| - | - | 350 | 3.6 µm | 8µm | no | Kendall et al (2011;2015) |
| UGC3825 | 110 | 160 | Hα | 0.4µm | yes | Peterken et al (2019) |
| Milky Way | - | 140 | CO 1-0 | masers | yes | Vallée (2014) |
| - | - | 210 | CO 1-0 | HII regions | no | Hou & Han (2015) |
| - | - | 30 | CO 1-0 | masers | yes | Vallée (2016) |
| - | - | 400 | old stars | 70µm | no | Ragan et al (2018) |
| - | - | 100 | Cepheids | maser | no | Baba et al (2018) for Perseus |
| - | - | 100 | CO 1-0 | masers | yes | Vallée (2018) for Perseus |

Note: The resolution is that of the fitting function y(x). Tracer1 is often peaked on cold gas. Tracer2 is often peaked on stellar emission.

**Table 2. Recipe to focus on the detection of a physical offset between tracers in a spiral arm**

| Condition | Necessary or sufficient | Area |
|---|---|---|
| 1. the gas and stars must travel **through** a spiral arm (Note 1) | nec. | galaxy |
| 2. a density wave provides a shocked lane to create new stars | nec. | dust lane |
| 3. the two tracers are separated, reached by the gas at diff. time | nec. | wavelength |
| 4. the two tracers must be reddening-free | suff. | arm |
| 5. avoid HI as it is found in hot gas (dust lane) and cold gas (cloud) | suff. | arm |
| 6. pick resolution not too high (signal loss) nor too small (beam dilution) | nec. | telescope |
| 7. avoid ‘missing flux’ in interferometer (large-angular scale resolved out) | nec. | telescope |
| 8. pick simple fitting function y(x) (no negative lobes; see Note 2) | suff. | analysis |
| 9. the fitting function must cover an appropriate range in x (e.g., in arcsecs) | suff. | analysis |
| 10. the offset found must be larger than its error, at least by a factor of 3 | nec. | stats |

Note 1. Trajectories are not circular, passing through a spiral shock then flowing to some extent along the arm and then flowing to the inter-arm region (Roberts 1975; Dobbs & Pringle 2010).
Note 2. Simple functions y(x) are: Gaussian, box car, polynomial, etc. More complex ones are: Fourier transform, cross-correlation; auto-correlation; Q-index.

## Table 3. Detected physical offset (in parsecs) observed between tracers of spiral arms (excluding undetected offsets)

| Spiral Galaxy name | V (km/s) | **at R (note 1)** (kpc)(ref.) | typical linear offset (parsec)(ref.) | info (note 2) |
|---|---|---|---|---|
| NGC 0578 | 186 | at 9 (M9) | 970 (eq1) | incl. 51$^{o}$ |
| NGC 628 = M74 | 220 | at 3 (T8) | 213 (eq1) | - |
| same | | - | 125 (K18) | - |
| same | 061 | at 3 (E9) | 364 (eq1) | incl. 24$^{o}$ |
| NGC 0925 | 121 | at 5 (T8) | 376 (eq1) | - |
| NGC 1417 | 332 | at 8 (M9) | 728 (eq1) | incl. 52$^{o}$ |
| NGC 1566 | - | - | 440 (S18) | 5$^{o}$ at 5 kpc |
| NGC 2403 | 128 | at 2 (T8) | 95 (eq1) | - |
| NGC 2841 | 331 | at 5.5 (T8) | 440 (eq1) | - |
| NGC 3031 = M81 | 256 | at 4 (T8) | 74 (eq1) | - |
| NGC 3184 | 260 | at 4 (T8) | 194 (eq1) | - |
| same | 140 | at 2.5 (E9) | 426 (eq1) | incl. 17$^{o}$ |
| NGC 3351 | 210 | at 3 (T8) | 211 (eq1) | - |
| NGC 3521 | 242 | at 4 (T8) | 331 (eq1) | - |
| NGC 3621 | 144 | at 2 (T8) | 189 (eq1) | - |
| NGC 3627 | 204 | at 4 (T8) | 322 (eq1) | - |
| NGC 4254 | 140 | at 5 (P93; M9) | 900 (eq1) | incl. 42$^{o}$ |
| same | 114 | at 5 (E9) | 796 (eq1) | incl. 52$^{o}$ |
| NGC 4303 | 154 | at 4 (E9) | 628 (eq1) | incl. 27$^{O}$ |
| NGC 4321 | 231 | at 4 (E9) | 247 (eq1) | incl. 27$^{O}$ |
| NGC 4736 | 192 | at 1 (E9) | 447 (eq1) | incl. 35$^{O}$ |
| NGC 5055 | 209 | at 3 (T8) | 194 (eq1) | - |
| NGC 5194 = M51 | 242 | at 4 (T8) | 530 (eq1) | - |
| same | 200 | at 4 (L13) | 350 (L13) | - |
| same | - | - | 220 (Ca) | - |
| same | 228 | at 4 (E9) | 940 (eq1) | incl. 20$^{o}$ |
| NGC 5457 | 194 | at 2 (E9) | 201 (eq1) | incl. 18$^{O}$ |
| NGC 6946 | 201 | at 3 (T8) | 121 (eq1) | - |
| same | 200 | at 3 (E9) | 101 (eq1) | incl. 30$^{o}$ |
| NGC 7793 | 109 | at 1 (T8) | 83 (eq1) | - |
| UGC3825 | 190 | at 4.0 (P19) | 530 (eq.1) | - |
| Milky Way – GQ1 | 230 | at 5 (V17) | 340 (V16) | Galactic Quadrant I |
| Milky Way - GQ4 | 230 | at 5 (V17) | 360 (V16) | Galactic Quadrant IV |
| Median | 200 | at 4 | 326 | 24 galaxies |
| Mean and s.d.m. | 198 ±13 | | 370 ±50 | 24 galaxies |
| Mean (only T8) | 206 ±17 | | 241 ±37 | 14 galaxies |
| Mean (only E9) | 168 ±19 | | 461 ±93 | 9 galaxies |

Note 1: The relation is $V = \Omega_{gas} . R$

Note 2: the circular orbital velocity (V) has been corrected for a galactic inclination of this value.

References (ref.): Ca: Chandar et al (2017); E9: Egusa et al (2009); eq1: equation 1; K18: Kreckel et al (2018); L13: Louie et al (2013); M9: Martinez-Garcia et al (2009a); P19: Peterken et al (2019); P93: Phookun et al 1993; S18: Shabani et al (2018); T8: Tamburro et al 2008; V16: Vallée (2016); V17: Vallée (2017a); V18: Vallée (2018).

**Table 4. Starformation timescales (in million years) from the dust lane outward, in spiral galaxies (excluding undetected offsets)**

| Galaxy Name | $\Omega_{sp}$ (km/s/kpc) | mean time offset (Myr) | method y employed |
|---|---|---|---|
| NGC 0578 | 15 (M9) | 19.0 (M9) | Q index |
| NGC 628 = M74 | 26 (T8) | 1.5 (T8) | cross-correlation |
| same | 16 (E9) | 28 (E9) | offset method |
| NGC 0925 | 11 (T8) | 5.7 (T8) | cross-correlation |
| NGC 1417 | 35 (M9) | 14.0 (M9) | Q index |
| NGC 1566 | 23 (S18) | - | 2-point correlation |
| NGC 2403 | 30 (T8) | 1.4 (T8) | cross-correlation |
| NGC 2841 | 42 (T8) | 4.4 (T8) | cross-correlation |
| NGC 3031 = M81 | 27 (T8) | 0.5 (T8) | cross-correlation |
| NGC 3184 | 38 (T8) | 1.8 (T8) | cross-correlation |
| same | 51 (E9) | 34.1 (E9) | offset method |
| NGC 3351 | 38 (T8) | 2.2 (T8) | cross-correlation |
| NGC 3521 | 32 (T8) | 2.9 (T8) | cross-correlation |
| NGC 3621 | 31 (T8) | 2.3 (T8) | cross-correlation |
| NGC 3627 | 25 (T8) | 3.1 (T8) | cross-correlation |
| NGC 4254 | 19 (P93) | 20.0 (M9) | Q index |
| same | 10 (E9) | 12.4 (E9) | offset method |
| NGC 4303 | 24 (E9) | 10.8 (E9) | offset method |
| NGC 4321 | 31 (E9) | 2.3 (E9) | offset method |
| NGC 4736 | 166 (E9) | 17.2 (E9) | offset method |
| NGC 5055 | 20 (T8) | 1.3 (T8) | cross-correlation |
| NGC 5194 = M51 | 21 (T8) | 3.4 (T8) | cross-correlation |
| same | 30 (L13) | 4.3 (eq.1) | peak tracing |
| same | 40 (E9) | 13.8 (E9) | offset method |
| NGC 5457 | 72 (E9) | 4.0 (E9) | offset method |
| NGC 6946 | 36 (T8) | 1.3 (T8) | cross-correlation |
| same | 36 (E9) | 1.1 (E9) | offset method |
| NGC 7793 | 40 (T8) | 1.2 (T8) | cross-correlation |
| UGC3825 | 40 (P19) | 19 (P19) | offset method |
| Milky Way – GQ1 | 19 (V18) | 2.5 (eq.1) | tangent to each arm |
| Milky Way - GQ4 | 19 (V18) | 2.7 (eq.1) | tangent to each arm |
| Median | 31 | 4.0 | 24 galaxies |
| Mean and s.d.m. | 36 ±6 | 7.4 ±1.5 | 24 galaxies |
| Mean (only T8)) | 30 ± 3 | 2.4 ±0.4 | 14 galaxies |
| Mean (only E9) | 49 ±16 | 13.7 ±3.8 | 9 galaxies |

Note: equation 1 has been used, along with data from Table 1, to infer the relevant values.

References: Ca: Chandar et al (2017); E9: Egusa et al (2009); eq.1: equation 1; K18: Kreckel et al (2018); L13: Louie et al (2013); M9: Martinez-Garcia et al (2009a); P19: Peterken et al 2019; P93: Phookun et al 1993; S18: Shabani et al (2018); T8: Tamburro et al 2008; V16: Vallée (2016); V17: Vallée (2017a); V18: Vallée (2018).

## Table 5. Angular and linear speeds, co-rotation radius, in spiral galaxies (excluding undetected offsets)

| Galaxy Name | $\Omega_{gas}$ (km/s/kpc) | $\Omega_{gas}$ - $\Omega_{sp}$ (km/s/kpc) | $v_{gas} - v_{sp}$ (km/s) | $R_{coro}$ (kpc) | |
|---|---|---|---|---|---|
| NGC 0578 | 20.7 | 5.7 | 51.3 | 12.4 | |
| NGC 628 = M74 | 46.8 | 25.8 | 77.4 | 6.7 | |
| NGC 0925 | 24.2 | 13.2 | 66.0 | 11.0 | |
| NGC 1417 | 41.5 | 6.5 | 52.0 | 9.5 | |
| NGC 1566 | - | - | - | - | |
| NGC 2403 | 64.0 | 34.0 | 68.0 | 4.3 | |
| NGC 2841 | 60.2 | 18.2 | 100.1 | 7.9 | |
| NGC 3031 = M81 | 64.0 | 37.0 | 148.0 | 9.5 | |
| NGC 3184 | 60.5 | 16.0 | 52.0 | 4.5 | |
| NGC 3351 | 70.0 | 32.0 | 96.0 | 5.5 | |
| NGC 3521 | 60.5 | 28.5 | 114.0 | 7.6 | |
| NGC 3621 | 72.0 | 41.0 | 82.0 | 4.6 | |
| NGC 3627 | 51.0 | 26.0 | 104.0 | 8.2 | |
| NGC 4254 | 25.0 | 10.5 | 52.5 | 8.8 | |
| NGC 4303 | 38.5 | 14.5 | 58.0 | 6.4 | |
| NGC 4321 | 57.8 | 26.8 | 107.2 | 7.5 | |
| NGC 4736 | 192.0 | 26.0 | 26.0 | 1.2 | |
| NGC 5055 | 69.7 | 49.7 | 149.0 | 10.4 | |
| NGC 5194 = M51 | 55.8 | 25.5 | 102.0 | 7.4 | |
| NGC 5457 | 97.0 | 25.0 | 50.0 | 2.7 | |
| NGC 6946 | 67.0 | 31.0 | 93.0 | 5.6 | |
| NGC 7793 | 109.0 | 69.0 | 69.0 | 2.7 | |
| UGC3825 | 47.5 | 7.5 | 30.0 | 4.8 | |
| Milky Way | 46.0 | 27.0 | 135.0 | 12.1 | |
| Median | 60.2 | 26.0 | 77.4 | 7.4 | 23 galaxies |
| Mean | 62.6 ±7.3 | 25.9 ±3.0 | 81.8 ±7.2 | 7.0 ±0.6 | 23 galaxies |

**Note:** $\Omega_{gas}$ in Column 2 is computed at the galactic radius given in col. 3 of Table 3. All input data from Tables 3 and 4. The shock speed (4th, multiplied by the gal; column) is computed from the previous column (angular speed difference) and multiplied by the galactic radius given in col.3 of Table 3.

**Figure Captions**

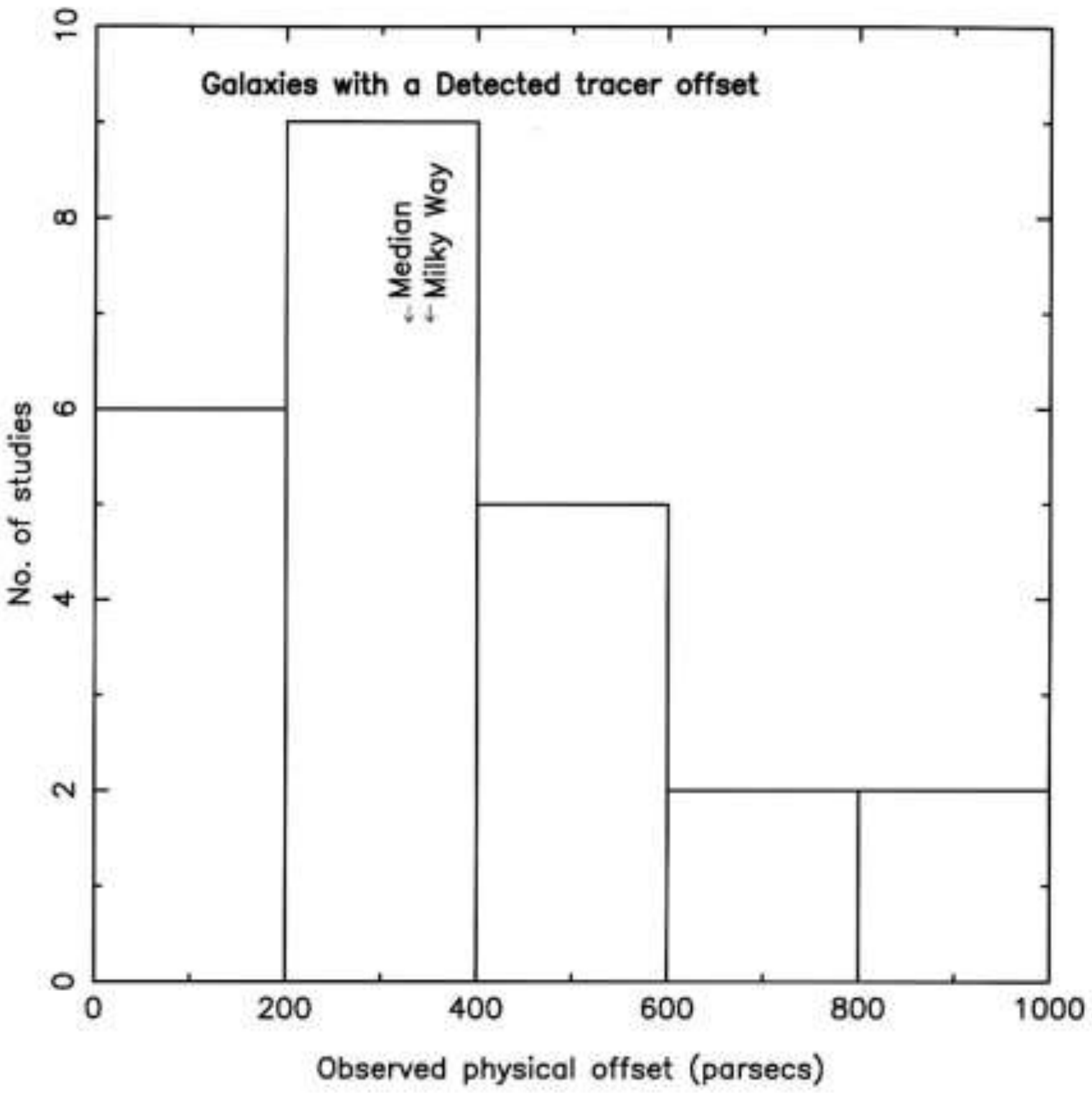

Figure 1. Galaxies with a detected offset. Histogram of the observed physical offsets from the dust lane, into the spiral arms, of several nearby galaxies. Data from Table 3. The median offset for these galaxies, and the Milky Way's offset, are shown. The various methods employed by others in this collection are different, as explained in the text, and a given method's bias is often countered by another method's bias, still revealing reasonable statistical means, at the price of an enlarged broadness of the values.

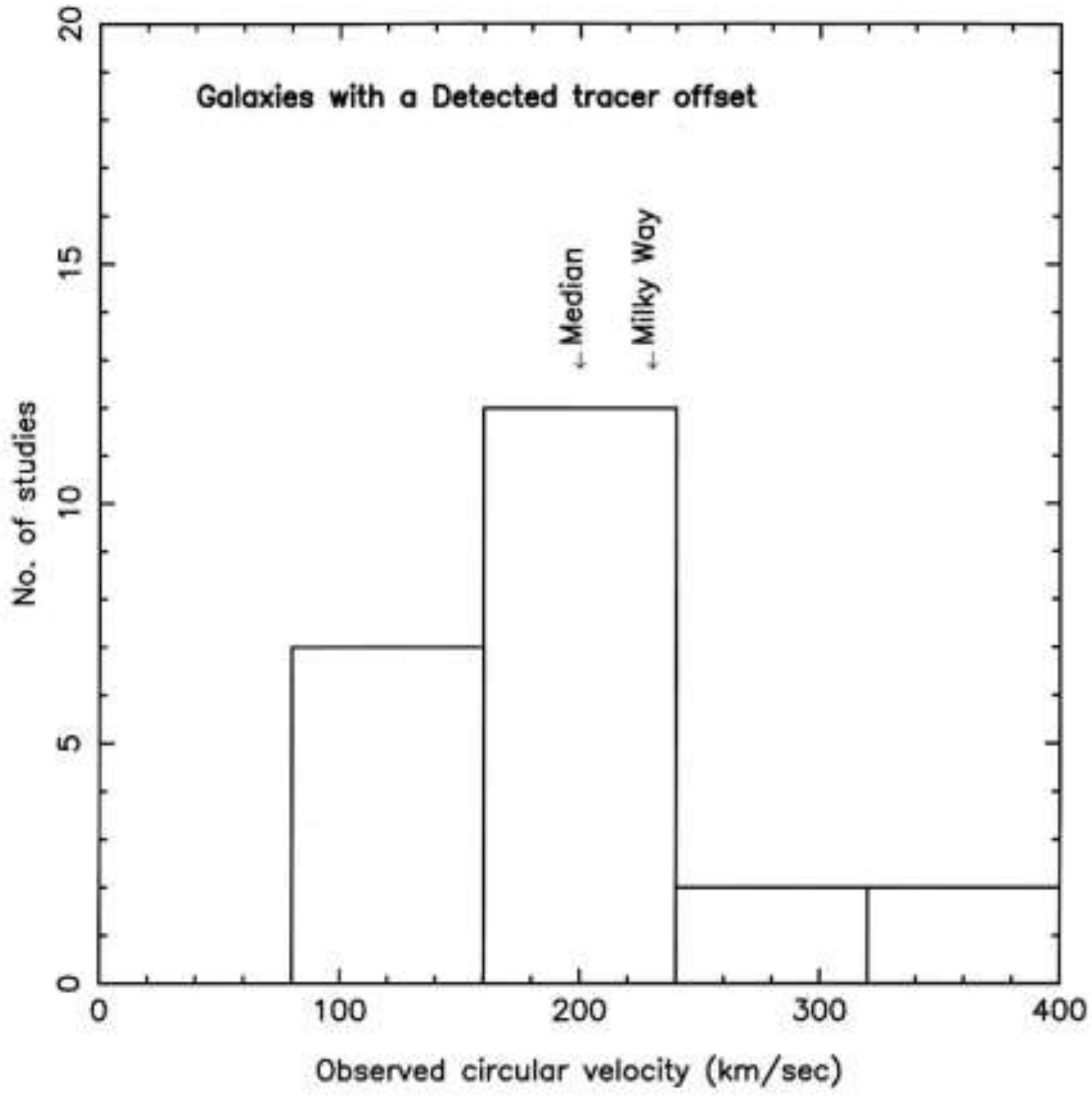

Figure 2. Galaxies with a detected offset. Histogram of the observed orbital circular velocity of the gas and stars around their galactic centers, at the galactic radius where the mean offset of tracers were observed. Data from Table 3. The median circular orbital velocity for these galaxies, and the one for the Milky Way, are shown. The various methods employed by others in this collection are different, as explained in the text, and a given method's bias is often countered by another method's bias, still revealing reasonable statistical means, at the price of an enlarged broadness of the values.

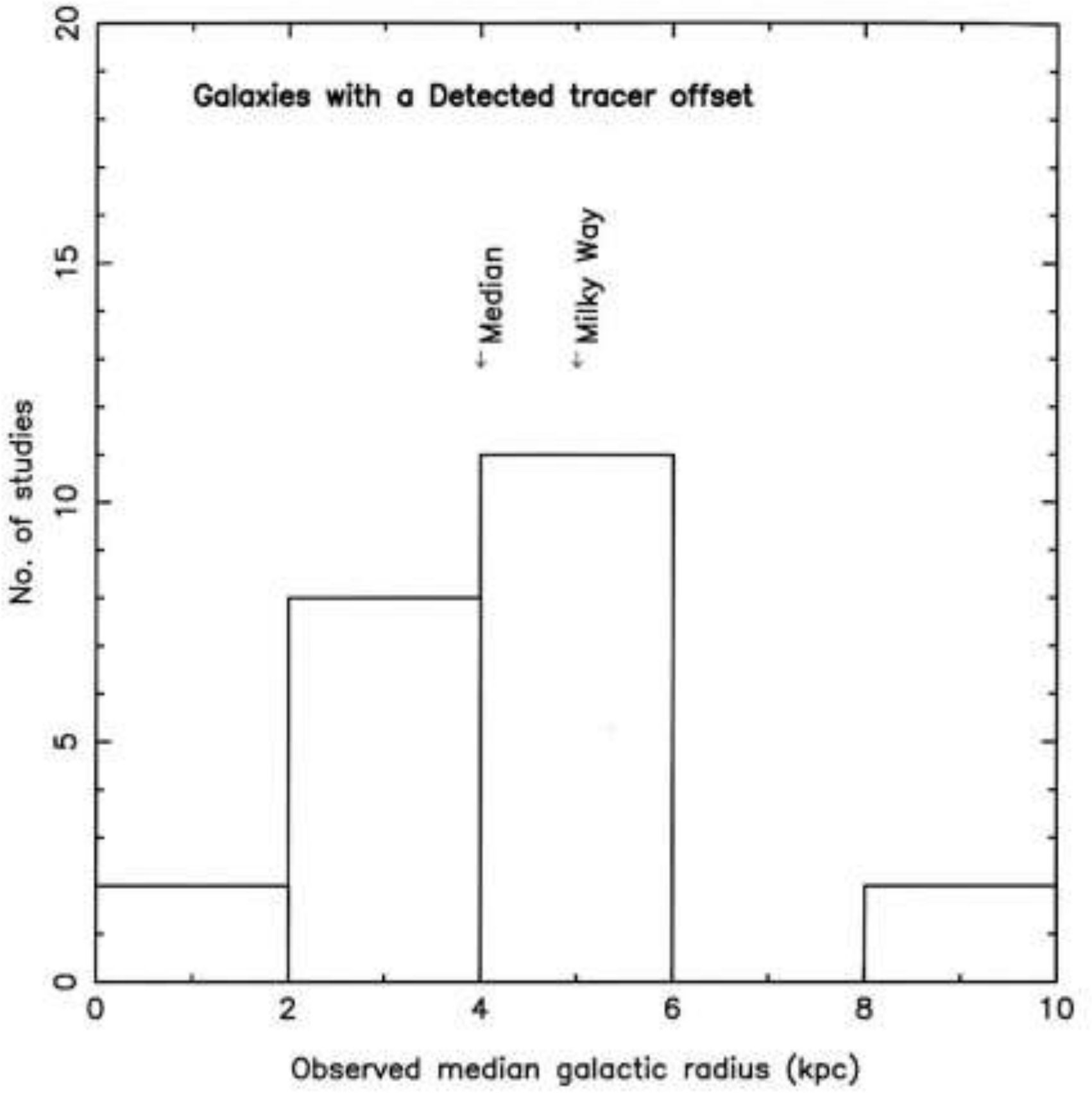

Figure 3. Galaxies with a detected offset. Histogram of the median galactic radius where the tracer offsets were measured. Data from Table 3. The median value for these galaxies, and the one for the Milky Way, are shown. The various methods employed by others in this collection are different, as explained in the text, and a given method's bias is often countered by another method's bias, still revealing reasonable statistical means, at the price of an enlarged broadness of the values.

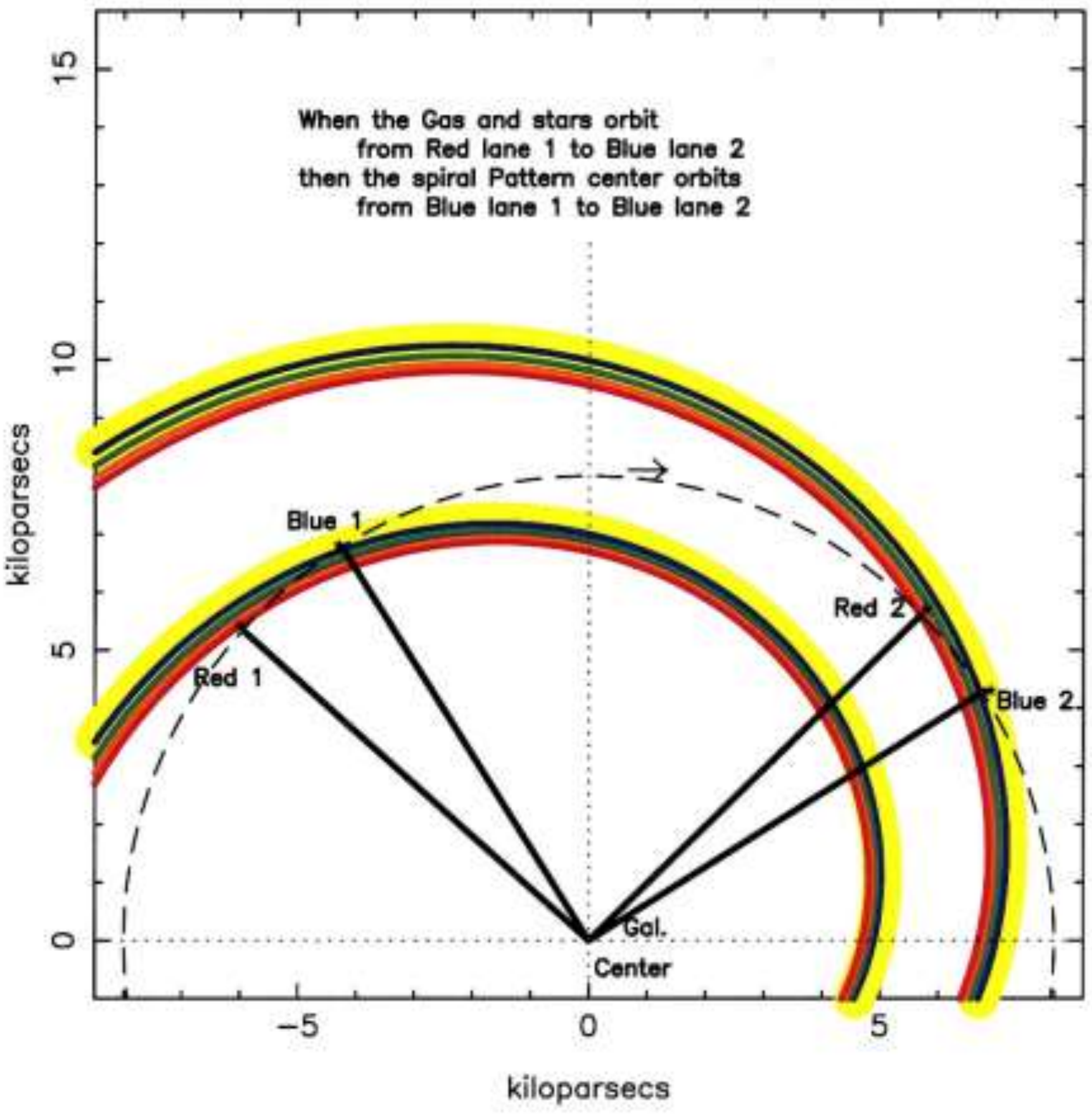

Figure 4. Sketch of the orbital trajectory of gas and stars over time, and the trajectory of the density wave spiral pattern – following the model of Vallée (2017c).

Color coding: the inner arm is the hot dust lane (red), followed by the lane of cold dust and radio masers (orange). A lane of observable young stars follow (green). The extended diffuse CO 1-0 molecules and the extended diffuse region of old stars, both found throughout the arm (yellow), have a peak near the arm mid-arm (black).

We follow a collapsing clump orbiting clockwise, starting from Red lane 1 (Red 1) at left to become eventually an old star located in Blue lane 2 (Blue 2) at right, then during that time scale the spiral pattern orbited clockwise from the Blue lane 1 (Blue 1) at left to reach the Blue lane 2 (Blue 2) at right.

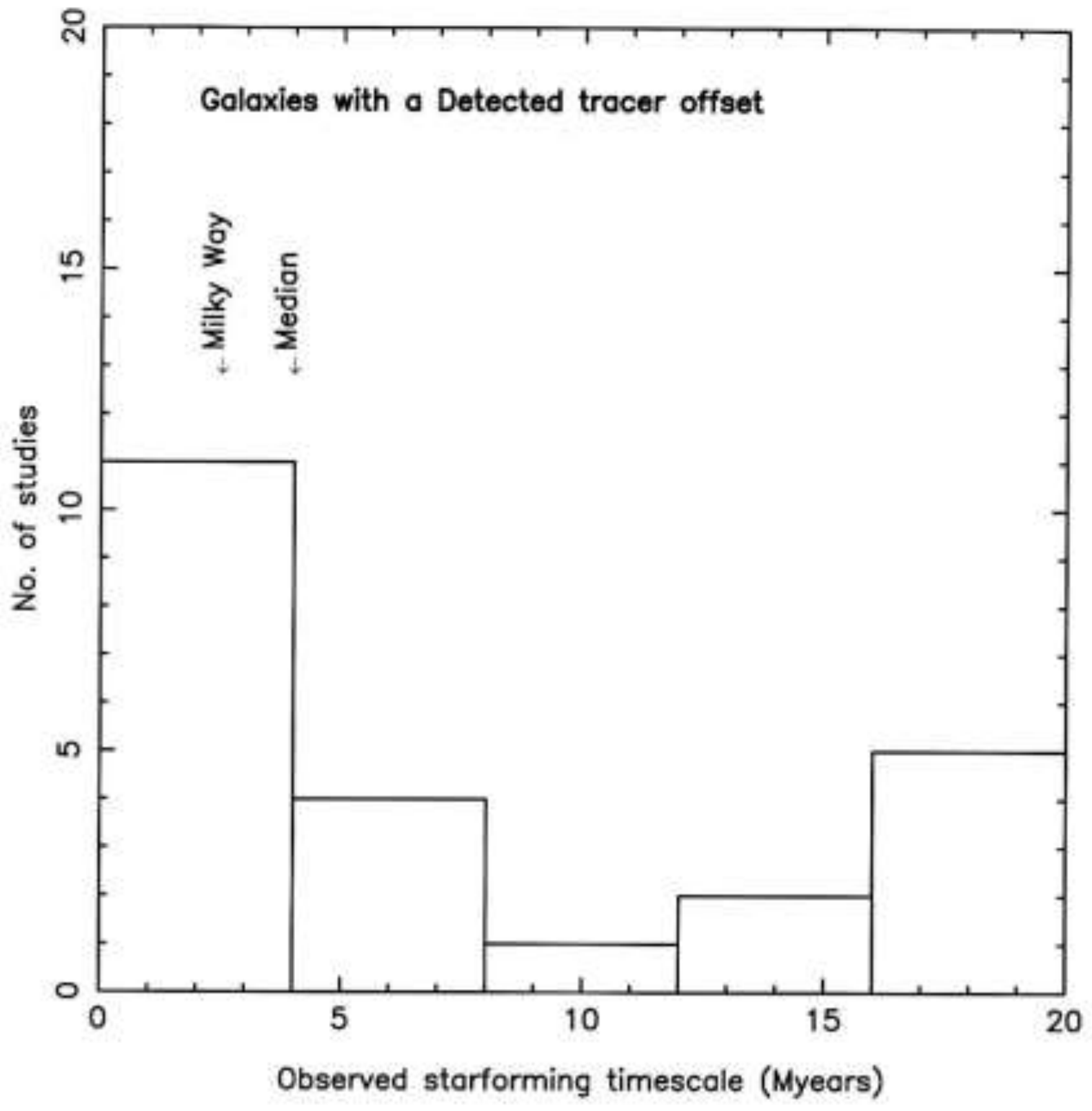

Figure 5. Galaxies with a detected offset. Histogram of the deduced starforming timescale (from new stars to old stars), in a spiral arm. Data from Table 4. The median value for these galaxies, and the one for the Milky Way, are shown. The various methods employed by others in this collection are different, as explained in the text, and a given method's bias is often countered by another method's bias, still revealing reasonable statistical means, at the price of an enlarged broadness of the values.

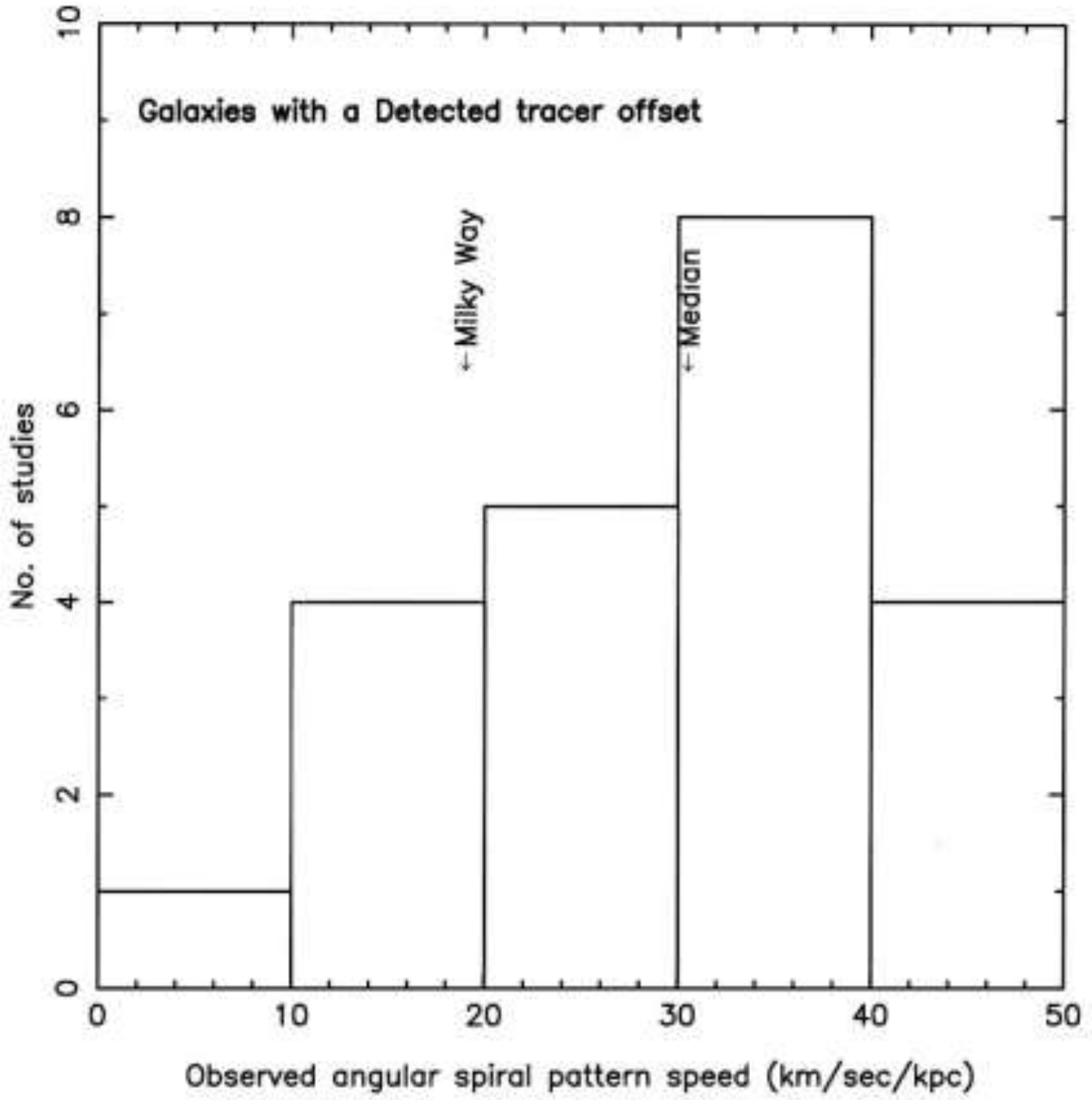

Figure 6. Galaxies with a detected offset. Histogram of the deduced angular spiral pattern speed around their galactic centers. Data from Table 4. The median value for these galaxies, and the one for the Milky Way, are shown. The various methods employed by others in this collection are different, as explained in the text, and a given method's bias is often countered by another method's bias, still revealing reasonable statistical means, at the price of an enlarged broadness of the values.

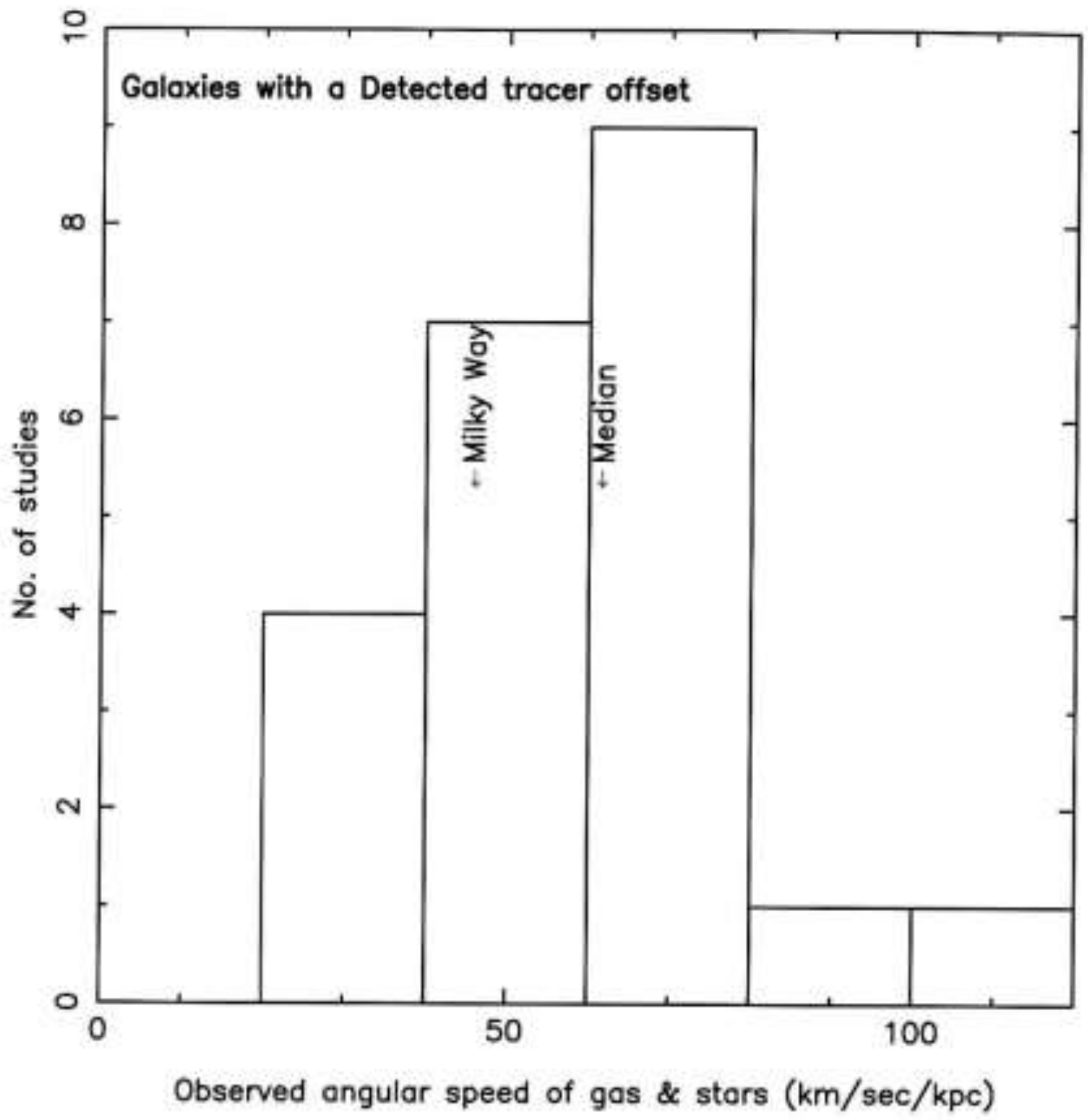

Figure 7. Galaxies with a detected offset. An histogram is shown of the angular orbital speed of the gas and stars. Data from Table 5. The median value for these galaxies, and the one for the Milky Way, are shown. The various methods employed by others in this collection are different, as explained in the text, and a given method's bias is often countered by another method's bias, still revealing reasonable statistical means, at the price of an enlarged broadness of the values.

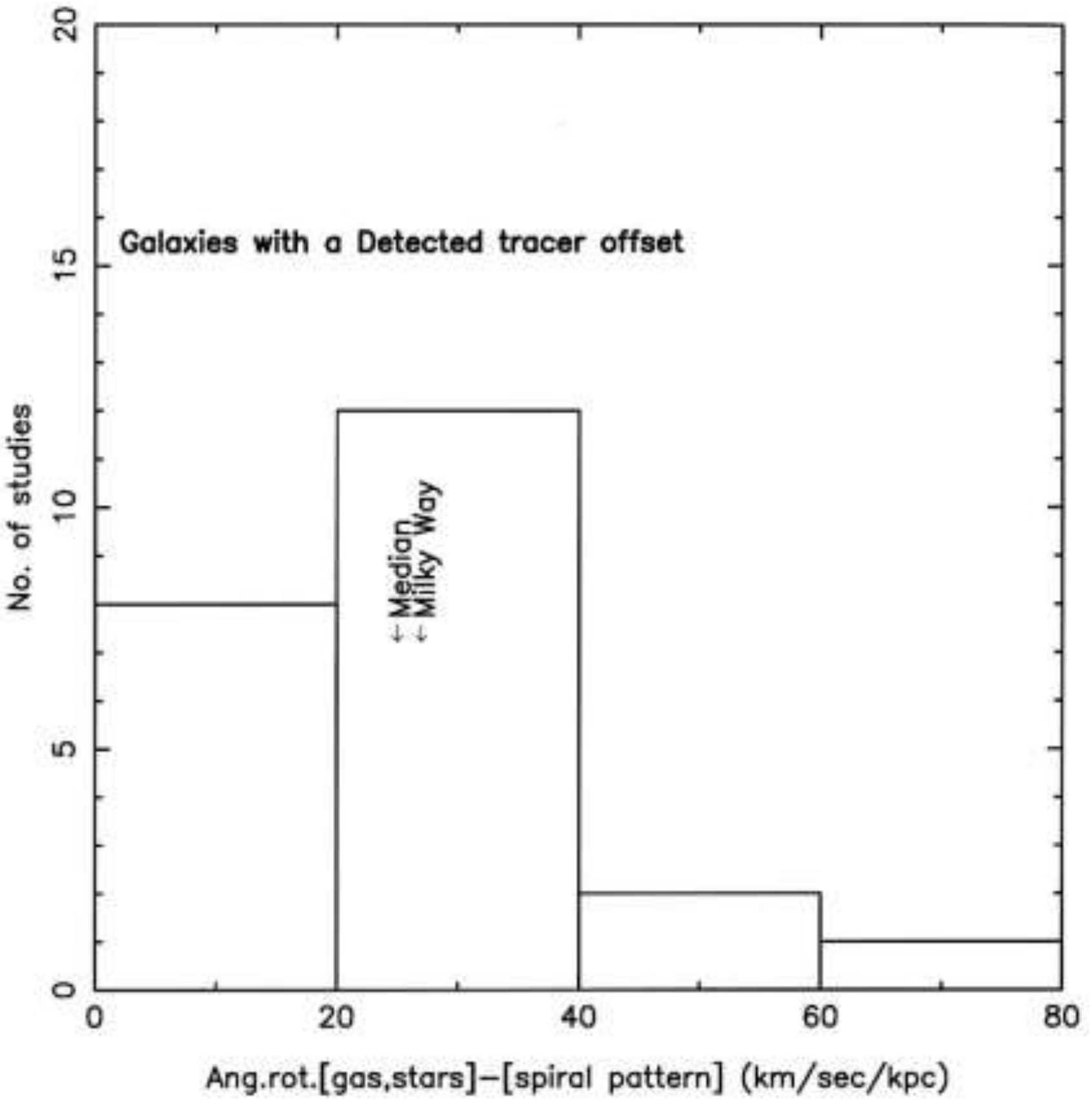

Figure 8. Galaxies with a detected offset. An histogram is shown of the difference in angular speed (gas and stars minus spiral pattern). Data from Table 5. The median value for these galaxies, and the one for the Milky Way, are shown. The various methods employed by others in this collection are different, as explained in the text, and a given method's bias is often countered by another method's bias, still revealing reasonable statistical means, at the price of an enlarged broadness of the values.

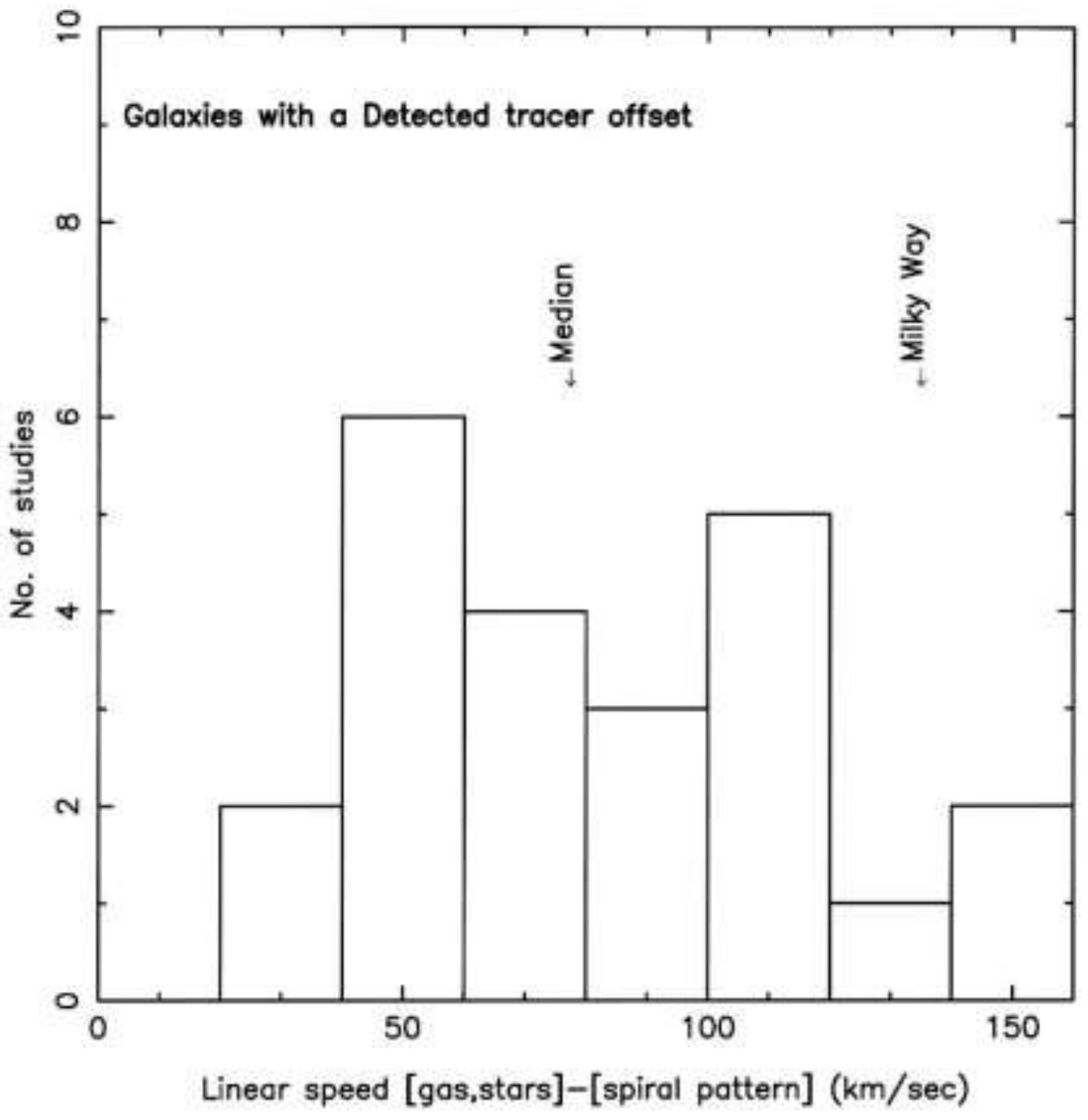

Figure 9._ Galaxies with a detected offset. An histogram of the shock speed is shown. Data from Table 5. The median value for these galaxies, and the one for the Milky Way, are shown. The various methods employed by others in this collection are different, as explained in the text, and a given method's bias is often countered by another method's bias, still revealing reasonable statistical means, at the price of an enlarged broadness of the values.

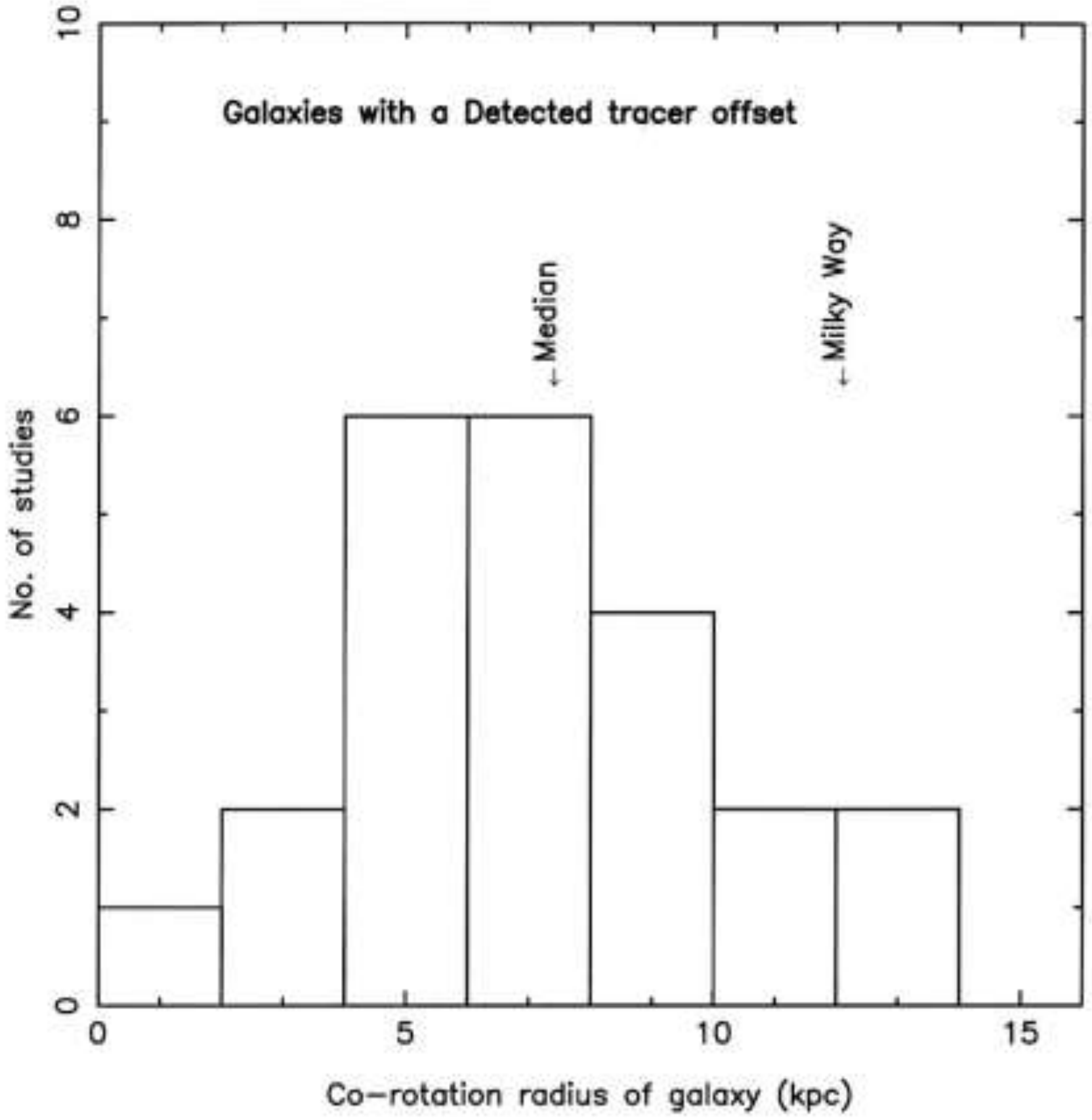

Figure 10. Galaxies with a detected offset. An histogram of the galactic co-rotation radius is shown. Data from Table 5. The median value for these galaxies, and the one for the  Milky Way, are shown. The various methods employed by others in this collection are different, as explained in the text, and a given method's bias is often countered by another method's bias, still revealing reasonable statistical means, at the price of an enlarged broadness of the values.

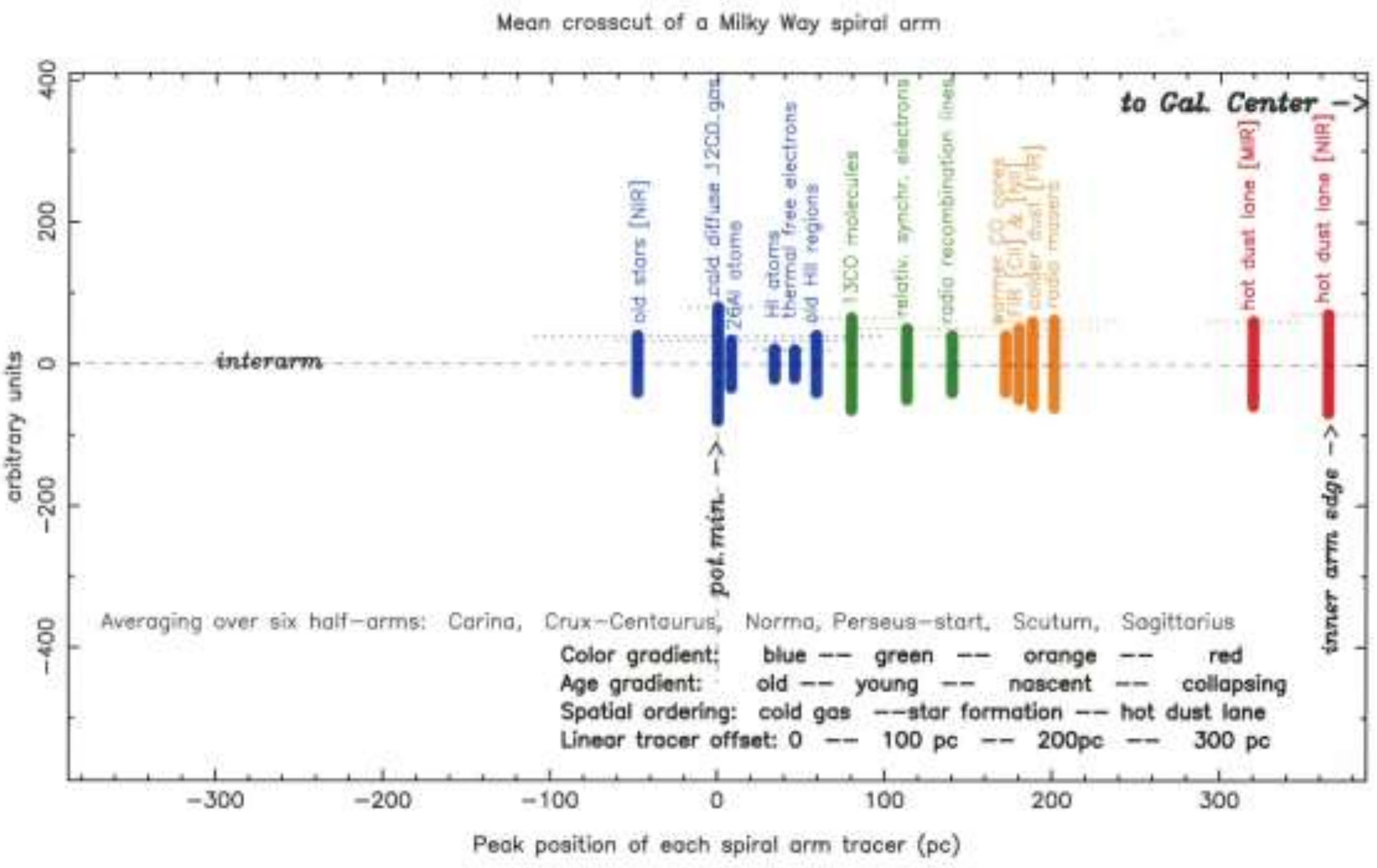

Figure 11. Outline of a spiral arm in the Milky Way.

Top of figure: new stars collapse in the red zone (at right), then proceed to the radio-detectable nascent stars (orange zone), then to the optically visible young stars (green zone), and finally to the cold zone at mid-arm (blue zone). This sketch shows a crosscut of a spiral arm, with the direction of the Galactic Center to the right. The broad diffuse cold CO 1-0 molecules, with a peak near the mid-arm (0 pc), is near the location of the 'potential minimum' of the density-wave theory. The inner arm edge is at the location of the 'shock' in the density-wave theory. The interarm location is shown at left.

Bottom of figure: different gradients are shown - color gradient, age gradient, spatial ordering, and linear offset of tracers. In the density wave theory, the 'shocked lane' is at the right of this plot, while the location of the 'potential minimum' (end of star formation) is at the center of this plot.

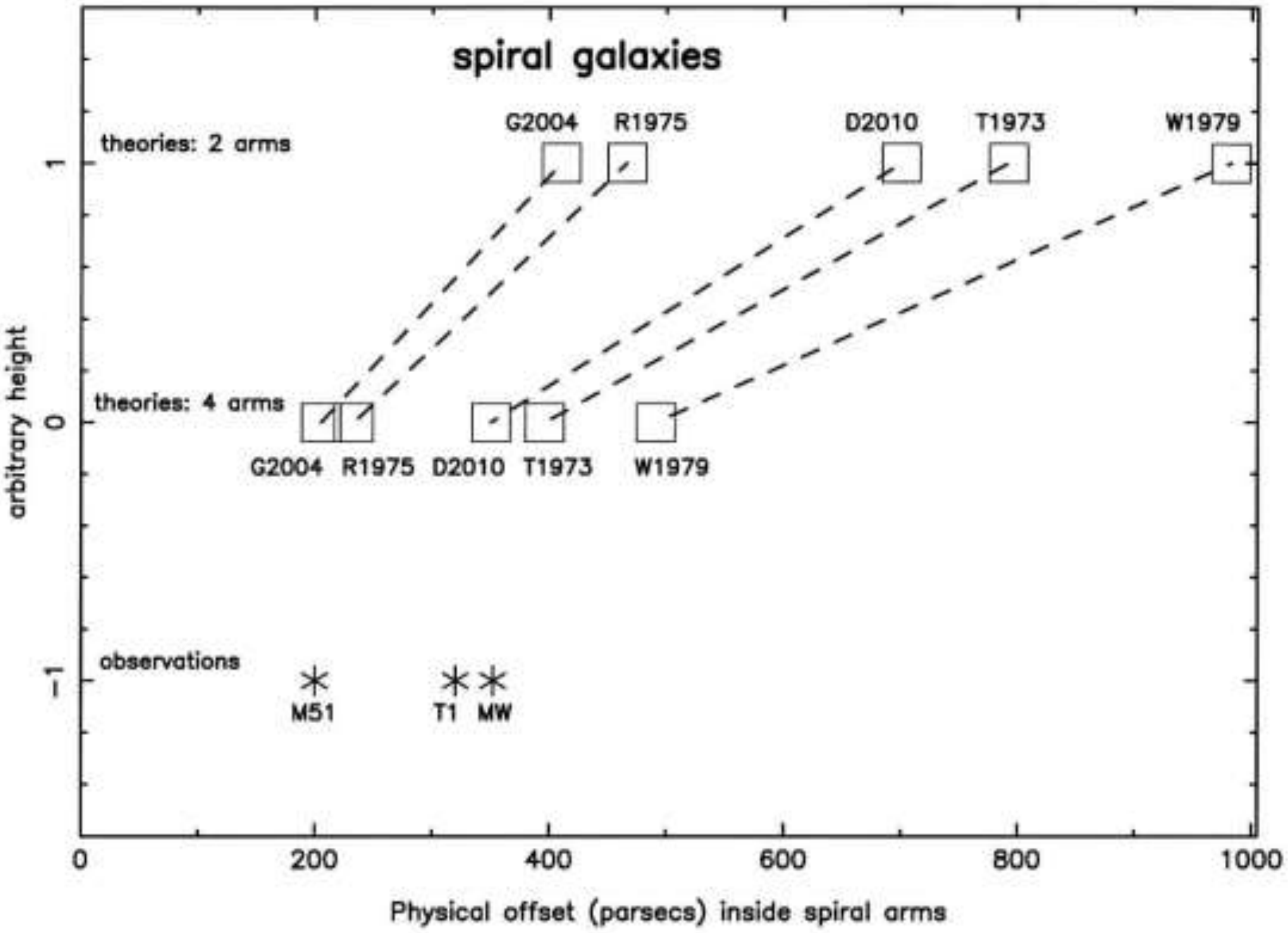

Figure 12. Predicted and observed values for the physical offset for the star-forming zone (horizontal axis). The predicted offset is the distance between the 'shocked lane' and the location of the 'potential minimum' of the density wave. The observed offset is the separation between an arm tracer in the shocked lane and an arm tracer at the end of the recent starformation.

The vertical axis y is in arbitrary units, allowing for the observations (at y= -1), the 4-arm predictions (at y=0) and the 2-arm predictions (at y= +1). The predicted value are shown as open squares. Predictions were scaled to a galactic radius of 4 kpc, separately for a 4-arm or a 2-arm model.

At the bottom, some of the observed values are shown as asterisks. Observational references are: M51 (Chandar et al 2017); MW (Vallée 2016); T1 (median of values in table 1 in this paper).

Theoretical references: G2004 (Gittens & Clarke 2004); R1975 (Roberts 1975); D2010 (Dobbs & Pringle 2010); T1973 (Tosa 1973); W1979 (Wielen 1979).